\documentclass[11pt]{myclass}
\usepackage{algorithm}
\usepackage[noend]{algorithmic}
\usepackage{listings}
\usepackage{amssymb,amsmath,amsfonts,pdfpages,color, url}
\usepackage{booktabs}
\usepackage{wrapfig}
\usepackage{multirow}

\newtheorem{example}{Example}[section]

\newcommand{\R}{\ensuremath{\mathbb{R}}}
\newcommand{\cR}{\ensuremath{\mathcal{R}}}

\newcommand{\cH}{\ensuremath{\mathcal{H}}}
\newcommand{\cA}{\ensuremath{\mathcal{A}}}

\newcommand{\cQ}{\ensuremath{\mathcal{Q}}}
\newcommand{\W}{\ensuremath{\mathbb{W}}}
\newcommand{\vol}{\ensuremath{{\rm Vol}}}
\newcommand{\kde}{\ensuremath{{\textsc{kde}}}}

\def\argmin{\mathop{\rm argmin}}

\def\dim{\mathop{\rm dim_{VC}}}
\def\E{\mathop{\mathbb{E}}}

\newcommand{\symdiff}{\mathbin{\triangle}}
\newcommand{\eps}{\varepsilon}

\begin{document}

\title{Dimension-Independent Kernel $\varepsilon$-Covers}

\author{Jeff M. Phillips and Hasan Pourmahmood-Aghababa \\ University of Utah \\ \texttt{jeffp@cs.utah.edu} and \texttt{h.pourmahmoodaghababa@utah.edu}}

\maketitle

\begin{abstract}
We introduce the notion of an $\eps$-cover for a kernel range space.  A kernel range space concerns a set of points $X \subset \R^d$ and the space of all queries by a fixed kernel (e.g., a Gaussian kernel $K(p,\cdot) = \exp(-\|p-\cdot\|^2)$, where $p \in \R^d$).  For a point set $X$ of size $n$, a query returns a vector of values $R_p \in \R^n$, where the $i$th coordinate $(R_p)_i = K(p,x_i)$ for $x_i \in X$.  An $\eps$-cover is a subset of points $Q \subset \R^d$ so for any $p \in \R^d$ that $\frac{1}{n} \|R_p - R_q\|_1\leq \eps$ for some $q \in Q$.  This is a smooth analog of Haussler's notion of $\eps$-covers for combinatorial range spaces (e.g., defined by subsets of points within a ball query) where the resulting vectors $R_p$ are in $\{0,1\}^n$ instead of $[0,1]^n$.  The kernel versions of these range spaces show up in data analysis tasks where the coordinates may be uncertain or imprecise, and hence one wishes to add some flexibility in the notion of inside and outside of a query range.  

Our main result is that, unlike combinatorial range spaces, the size of kernel $\eps$-covers is independent of the input size $n$ and dimension $d$.  We obtain a bound of $2^{\tilde O(1/\eps^2)}$, where $\tilde{O}(f(1/\eps))$ hides log factors in $(1/\eps)$ 
that can depend on the kernel.  This implies that by relaxing the notion of boundaries in range queries, eventually the curse of dimensionality disappears, and may help explain the success of machine learning in very high-dimensions.  
We also complement this result with a lower bound of almost $(1/\eps)^{\Omega(1/\eps)}$, showing the exponential dependence on $1/\eps$ is necessary.
\end{abstract}

\keywordlist{$\eps$-Cover, $\eps$-Sample, Kernel range space, Dimensionality reduction.}

\let\thefootnote\relax\footnotetext{Jeff Phillips thanks his support from NSF CCF-2115677, CCF-1350888, and IIS-1816149.  We thank an anonymous reviewer of an earlier version of the paper for suggesting the argument based on Radamacher complexity for reproducing kernels.  We also thank Nico Spronk for providing the proof of Lemma \ref{lem: sphere intersection-proof} through email communication.}

\newpage
\setcounter{page}{1}

\section{Introduction}

Given a data set $X$ a \emph{range space} $(X,\cR)$ is the collection of possible ways that set $X$ can be queried; $\cR$ is a set of subsets of $X$, often defined by intersection with a type of geometric shape.  
For a data structure, ranges specify the shape of any range query~\cite{agarwal2017range}.  
For machine learning, ranges categorize the function class of possible classifiers~\cite{vapnik1991principles}.  
For spatial scan statistics, ranges restrict the family of regions which might form an anomalous hotspot~\cite{kulldorff1997spatial}.  

In each of these cases, it is common to allow $\eps |X|$ additive error when considering the results of these queries.  In that context, an \emph{$\eps$-cover}, which is an instance of a cover in a metric space, and Haussler~\cite{haussler1995sphere} among others studied it for a particular class of metric spaces, is an important concept; it is a subset $\cQ$ of all possible subsets in the collection $(X,\cR)$ so that for any range $R \in (X,\cR)$ there exists some set $Q \in \cQ$ so that the symmetric difference $|Q \triangle R| \leq \eps|X|$.  
In particular, if one allows $\eps|X|$ error, then one only needs to consider each of the above listed data analysis challenges with respect to the $\eps$-cover $\cQ$, not the full collection of possible subsets.  

Haussler introduced and bounded the size of $\eps$-covers for range spaces with bounded VC-dimension~\cite{haussler1995sphere}.  In particular, if the VC-dimension $\nu$ is bounded, then there exist $\eps$-covers of size $O(1/\eps^\nu)$ and $\Omega(1/\eps^\nu)$ may be needed.   
Consider the common range spaces for $X \subset \R^d$ like half-space, ball, and fixed radius ball range spaces, each has VC-dimension $\nu = d+1$.  However, Haussler's lower bound does not apply to these geometric range spaces specifically.  Nevertheless, we supply a lower bound of $\Omega((1/\eps)^{d^{1-o(1)}})$ for them in Section \ref{app:lower-bound-eps-cover}.

In this paper, we consider how this changes when we consider kernelized versions of these objects; that is where ranges are defined by kernels, like Gaussian kernels $K(x,q) = \exp(-\|x-q\|^2)$.  Indeed, kernel SVM is a common way to build non-linear classifiers~\cite{scholkopf2018learning}, and kernelized versions of data structures queries~\cite{charikar2017hashing,charikar2020kernel,karppa2022deann} and scan statistics~\cite{han2019kernel,fitzpatrick2021support,herlands2018gaussian} are also common.  
Partially motivated by these cases, the complexity of kernel range spaces have also been studied, and in particular samples for density approximation.  These \emph{$\eps$-KDE-samples} are subsets $S \subset X$ so for every query $p \in \R^d$ that 
\[
\left| \frac{1}{|X|} \sum_{x \in X} K(x,p) - \frac{1}{|S|}\sum_{s \in S} K(s,p) \right|
=
\left|\kde_X(p) - \kde_S(p)\right|
\leq \eps.
\]
While for positive and symmetric kernels, a bound of $O(d/\eps^2)$ for such an $\eps$-KDE-sample can be derived using bounds for ball range spaces~\cite{joshi2011comparing}, more remarkably, for reproducing kernels, only size $O(1/\eps^2)$ is needed~\cite{lopez2015towards,LLB2015,bach2012equivalence}, that is with no dependence on $d$.

We tackle whether a similar result, with no dependence on $n$ or $d$ is possible for an $\eps$-cover of a kernel range space.  In particular, a kernel range space $(X,K)$ is defined by a set of input points $X \subset \R^d$, and a fixed kernel $K$, e.g., Gaussians of the form $K(p,\cdot) = \exp(-\|p-\cdot\|^2)$.  In this setting, any \emph{range} in the kernel range space is defined by a point $p \in \R^d$, and reports a signature vector $R_p^X = (K(p,x_1), K(p,x_2), \ldots, K(p,x_n)) \in \R^n$, which has a scalar value $K(p,x_i)$ for each $x_i \in X$;  we consider when $K(p,x_i) \in [0,1]$.  This generalizes the notion of a set, where these signatures are bit-vectors from $\{0,1\}^n$ instead of $[0,1]^n$.  An \emph{$\eps$-cover of a kernel range space $(X,K)$} is then a set of kernel ranges $K(q,\cdot)$, defined by a set of points $Q \subset \R^d$, so for \emph{any} query point $p \in \R^d$ there exists a $q \in Q$ so that 
\[
\frac{1}{|X|} \sum_{x \in X} |K(p,x) - K(q,x)| = \frac{1}{|X|}\|R_p^X - R_q^X\|_1 \leq \eps.
\]

This generalizes the notion of $\eps$-cover of Haussler to function values.  Notice that if we instead placed the absolute values outside the sum, this becomes trivial since one can simply choose $1/\eps$ points $Q$ where $\kde_X(q_i) = (i-1/2)/\eps$ for $i = 1, \ldots, 1/\eps$.

\subparagraph*{Our Results.}
Our main result is that $\eps$-covers for kernel range spaces have size complexity independent of $n$ and $d$.  Thus for constant error (e.g., $\eps = 0.01$ for $1\%$ error), the size of the $\eps$-cover is constant; that is, to evaluate these functions up to a fixed error, one only needs to pre-compute or consider evaluating a fixed number of kernel range queries.  
In particular, we show that the size of $\eps$-covers are at most $2^{\tilde O(1/\eps^2)}$; where $\tilde O(f(1/\eps))$ hides polylogarithmic factors in $1/\eps$.    
This bound works for a large class of kernels we call ``standard'' and includes Gaussian, Laplace, truncated Gaussian, triangle, Epanechnikov, quartic, and triweight.  
Moreover, we show that this $(1/\eps)^{\mathsf{poly}(1/\eps)}$ is necessary.  In particular, for Gaussian kernels we provide a construction that requires an $\eps$-cover of size $(\frac{1}{\eps})^{\Omega(1/\eps^\lambda)}$ for any $\lambda \in (0,1)$ in $\R^{d'}$ with $d' = \Omega(1/\eps^\lambda)$.  

When viewed in comparison to the $\eps$-cover size bound for traditional range spaces, e.g. for half-spaces or balls, where the size grows exponentially in $d$ (see discussion in Section \ref{app:lower-bound-eps-cover}), we believe this result is quite surprising.  Almost all learning or data structure bounds, even approximate ones, have exponential dependence on $d$ in the number of queries considered.  
However, this result shows that if one relaxes the boundary of the query, that is there is not a hard or combinatorial cut-off separating ``in'' the query or ``not in'' the query, then this exponential dependence and curse of dimensionality (eventually) disappears.

\subparagraph*{Overview of Techniques. }
One may think (as we initially hoped) that this $\eps$-cover result is a not-too-hard consequence of the dimension-independent bounds for $\eps$-KDE-samples.  However, these results seem to provide the wrong sorts of guarantees; they would work if the definition of the $\eps$-cover had the absolute values outside the sum.  Moreover, both constructions for $\eps$-KDE-samples rely on properties of reproducing kernels, namely that the kernel density estimate $\kde_X$ can be viewed as a mean in a reproducing kernel Hilbert space.  This quantity turns out to be easy to approximate with sub-gradient descent~\cite{LLB2015,bach2012equivalence} or sampling~\cite{lopez2015towards}.  However, the $\eps$-cover is a richer and more structured summary of a point set, and does not admit such simple analysis.  

Our approach at its core uses the simple idea that for a kernel with a bounded support (of value above $\eps$), one can place a grid around each data point with a gap of $\eps$ between grid points.  The union of all grid points is the $\eps$-cover.  Naively, this provides a bound of roughly $n (1/\eps)^d$ for $n = |X|$ points in $\R^d$.  
This paper shows how to preserve the correctness of this construction while reducing both $n$ and $d$ to only depend on $\eps$.  

Being able to remove the dependence on $n$ is perhaps not that surprising given the existence of $\eps$-KDE-samples and similar data reduction results.  However, this required some new adaptations on existing ideas as the direct invocation of $\eps$-KDE-samples does not work.  
We connect to $\eps$-samples of (traditional) range spaces, which we call \emph{semi-linked} to kernels, and their VC-dimension.  These semi-linked ranges are defined as the \emph{super-level sets of the difference of two kernel functions}.  A key insight is that these semi-linked range spaces allow us to calculate an intermediate object called an $\eps$-cover-sample, via a simple random sample, and this $\eps$-cover-sample can be converted into an $\eps$-cover. We show for Gaussian, triangle, Epanechnikov, quartic, and triweight kernels that this VC-dimension bound is $O(d^2)$.  So this reduction eliminates the dependence on size $n$, but adds dependence on dimension $d$.  

More surprising to the authors is that the dependence on the dimension can be eliminated.  The argument works by showing the existence of an embedding into dimension $m = O((1/\eps^2) \log n)$ where the measurement of all kernels on the input point set $X$ is preserved up to $\eps$ error.  This embedding is a result of invoking terminal JL~\cite{NN2019}.  It preserves the $R_p^X$ signatures for each point $p$, and means it is sufficient to create an $\eps$-cover in that $m$-dimensional space.  However, since the terminal JL embedding map is not invertible for points not in $X$, we require a new combinatorial covering argument to show that an $\eps/8$-cover in $\R^m$ is still an $\eps$-cover in the original $\R^d$. 
While this process eliminates the dependence on $d$, it increases it with respect to the number of points $n$.  

Iterating between these two approaches would reduce both $n$ and $d$, but would naively require $\log^*(nd)$ iterations, potentially inflating the error by that factor, and so not be independent of one of those two terms.  
Luckily, however, we can adapt a new inductive framework~\cite{csikos2022optimal} for analyzing such iterative reduction processes, and we show a complete elimination of the dependence on $n$ and $d$.  
For positive definite kernels we apply a Rademacher complexity bound to calculate $\eps$-cover-samples independent of $d$; this does not immediately remove $\eps$-cover dependence on $d$, but does sidestep some of this iterative analysis.

For the lower bound, in low dimensions, the construction works like one may expect for fixed-radius balls, which when their radius is sufficiently large, act like half-spaces.  The size is trivially $\Omega(1/\eps)$ in $\R^1$, and as we add each dimension we add a point ``orthogonal'' to the existing dimensions.  The ranges we must cover is the cross-product of these distance intervals from points in each dimension, leading to a $(1/\eps)^d$ lower bound for fixed-radius balls.  However, interestingly, this construction stops working for kernels as we approach $1/\eps$ dimensions.  This is the result of both the curvature of the level sets and the decaying contribution with distance: properties implicit in data with full-dimensional noise.

\subparagraph*{Implications. }
This model and result is relevant in data analysis applications where a complete trust in data coordinates is rare (e.g., due to sensing noise), and it is common to have high dimensional data, and this sort of additive $\eps$-error is tolerated if not expected.  We hope this sheds a bit of light onto why learning in such high-dimensional spaces is not as challenging as traditional curse-of-dimensionality bounds may suggest.  That is, if one assumes sensing noise, adding more and more features (dimensions) does not always generate more implicit query complexity.  

For instance, our new definition of kernel $\eps$-cover is also the notion required to enumerate all possible ranges that could lead to a distinct solution in the case of approximate range searching  applications or noise-aware statistical modeling (e.g., for evaluating a kriging model~\cite{oliver1990kriging} or Nadaraya-Watson kernel regression~\cite{zheng2017coresets}) or in enumeration for approximate spatial scan statistics~\cite{matheny2016scalable}. 
Spatial scan statistics search (or ``scan'') \emph{all} combinatorial distinct ranges (up to $\eps$-error) to find one that maximizes some statistic on the data -- a candidate for an anomaly.  
Recently, Han \emph{et.al.}~\cite{han2019kernel} defined a kernel spatial scan statistic where the ranges are replaced with kernel queries, and the goal is still to approximately maximize a function of data over all such queries; they provide a solution in $d=2$.  Our results show that eventually, the size of the space required to search stops growing exponentially with dimension.  

Another line of work~\cite{AEC2020, CW2003, EFR2012} attempts to bound the number of local maximums of a Gaussian $\kde_X$ for $X \subset \R^d$ and $|X|=n$.  A lower bound is ${n \choose d} + n = \Omega(n^d)$ for $n,d \geq 2$~\cite{AEC2020}, but the upper bound is not known to be finite.  If one assumes finiteness, the best bound is $2^{d + {n \choose 2}}(5+3d)^n$.  If one only counts local maximum $p,q$ that have $\frac{1}{|X|} \|R_p^X - R_q^X\|_1 > \eps$ (i.e., are sufficiently distinct), then our result induces a bound of $O(2^{\tilde O(1/\eps^2)})$. 

Finally, this result induces the first dimension-independent bounds for $\eps$-KDE-samples~\cite{phillips2020near} for non-reproducing kernels including triangle, Epanechnikov, and truncated Gaussians.

\subsection{Connection to Uniform Glivenko-Cantelli Classes} \label{sec:Glivenko-Cantelli}
Starting with Vapnik and Chervonenkis~\cite{vapnik1971uniform}, numerous learning theorists, probabilists, and combinatorists have studied a strong and general notion of convergence of function approximation under sampling known as the \emph{uniform Glivenko-Cantelli} class.  
It concerns a class of functions $\mathcal{F}$ from a set $\mathcal{X}$ to $[0,1]$.  
Then let $\mathcal{P}$ be a probability measure over $\mathcal{X}$ so that any $f \in \mathcal{F}$ is $\mathcal{P}$-measurable over $\mathcal{X}$.  
Next we use $\mathcal{P}(f) = \int_{x \in \mathcal{X}} f(x) \mathsf{d} \mathcal{P}$ to denote the mean of $f(x)$ for $x \sim \mathcal{P}$. 
Now for an independent random sample $x_1, x_2, \ldots, x_m \sim \mathcal{P}$, let $\mathcal{P}_m(f) = \frac{1}{m} \sum_{i=1}^m f(x_i)$ be its approximation by the sample of size $m$.  

Dudley \emph{et.al.}~\cite{dudley1991uniform} defines that the family $\mathcal{F}$ is \emph{$\eps$-uniform Glivenko-Cantelli class} if
\[
\lim_{m \to \infty} \sup_{\mathcal{P}} 
 Pr[ 
 \sup_{k \geq m} \sup_{f \in \mathcal{F}} |\mathcal{P}_k(f) - \mathcal{P}(f)| > \eps
 ]= 0.  
\]
While bounded VC-dimension~\cite{vapnik1971uniform} implies that $\mathcal{F}$ is $\eps$-uniform Glivenko-Cantelli, it does not completely characterize this process.  

Alon \emph{et.al.} \cite{alon1997scale} showed that a variant of VC-dimension, called \emph{$V_\gamma$-dimension}, for any $\gamma > 0$, did characterize this form of convergence.  We say $\mathcal{F}$ \emph{$V_\gamma$-shatters} a set $A \subset \mathcal{X}$ if there exists a value $\alpha \in [0,1]$ such that for each $E \subseteq A$, there is another function $f_E \in \mathcal{F}$ so for all $x \in A \setminus E$ then $f_E(x) \leq \alpha - \gamma$ and for all $x \in E$ then $f_E(x) \geq \alpha + \gamma$.  The $V_\gamma$-dimension of $\mathcal{F}$ is the maximum cardinality $A \subseteq \mathcal{X}$ that is $V_\gamma$-shattered by $\mathcal{F}$.  
They showed that if the $V_\gamma$-dimension is finite, then $\mathcal{F}$ is $(b \gamma)$-uniform Glivenko-Cantelli for some constant $b \leq 48$.  Moreover, if the $V_\gamma$-dimension is not finite, then $\mathcal{F}$ is not $(2\gamma-\tau)$-uniform Glivenko-Cantelli for any $\tau > 0$.

The Glivenko-Cantelli criteria is intimately tied to $L_s$ $\eps$-covers via a result by Dudley \emph{et.al.} \cite{dudley1991uniform}.  An \emph{$L_s$ $\eps$-cover} is a set $F \subset \mathcal{F}$ so for any $f' \in \mathcal{F}$ it holds that some $f \in F$ satisfies $\|f - f'\|_s \leq \eps$, where $\|f - f'\|_s = (\int_{x \in \mathcal{X}} |f(x) - f'(x)|^s \mathsf{d}\mathcal{P})^{1/s}$ for $0 < s < \infty$, and $\|f - f'\|_\infty = \max_{x \in \mathcal{X}} |f(x) - f'(x)|$.  
Let $N_s(\eps,\mathcal{F},X)$ be the size of the smallest $L_s$ $\eps$-cover of $(X,\mathcal{F})$.   
Now let 
\[
H_m(\eps, \mathcal{F}) = \sup_{\substack{X \subset \mathcal{X} \\ |X|=m}} \log_2(N_s(\eps,\mathcal{F},X)).  
\]
They showed that $\mathcal{F}$ is Glivenko-Cantelli if and only if $\lim_{m \to \infty} H_m(\eps,\mathcal{F})/m = 0$.  Moreover, if $\eps > 0$ and $\lim_{m \to \infty} H_m(\eps,\mathcal{F})/m = 0$, then $\mathcal{F}$ is $(8 \eps)$-uniform Glivenko-Cantelli.  These results hold for any $s \in (0,\infty]$ in the definition of $\eps$-cover.

\subparagraph*{Application to Kernel Range Spaces. }
This paper studies a restrictive setting within this framework.  

First we consider $\mathcal{P}$ as the uniform probability measure on a fixed size-$n$ set $X$ with $\mathcal{X} = \mathbb{R}^d$.  Our results do allow $n$ to become arbitrarily large, and the sample complexity results do not depend on $n$, so it seems these approaches may extend naturally to continuous distributions $\mathcal{P}$.  However, we do not formalize this limiting case.  Moreover, we describe our results algorithmically, where the algorithms take a finite size $n$, and have run times which depend on $n$.  

Second, we consider a specific class of functions $\mathcal{F}_K$ where each $f_p(\cdot) = K(p,\cdot)$ takes the form of a kernel.  That is each $f_p \in \mathcal{F}_K$ is parameterized by a point $p \in \R^d$.  Then $\mathcal{P}(f_p) = \mathcal{P}(K(p,\cdot)) = \kde_X(p)$.  

Moreover, one can apply a Chernoff-Hoeffding bound on any one covering element (parameterized by $p$) with $O((1/\eps^2)\log(1/\delta))$  
samples to, with probability $1-\delta$, get $\eps$ error for any one evaluation point $p$. 
Then one can take a union bound over $N_1(\eps/2, \mathcal{F}_K, X)$ covering elements, and using triangle inequality, ensure that with $k_\eps = O((1/\eps^2)\log(\frac{N_1(\eps/2,\mathcal{F}_K,X)}{\delta})$ samples $S$, we have with probability at least $1-\delta$ that
\[
\sup_{f_p \in \mathcal{F}_K} |\mathcal{P}(f_p) - \mathcal{P}_m(f_p)| = \sup_{p \in \R^d} |\kde_X(p) - \kde_S(p)| \leq \eps.
\]
Classically, for binary functions families $f \in \mathcal{F}$ with $f(x) \in \{0,1\}$, this argument also works using $s = \infty$ \cite{vapnik1971uniform}; and in this binary setting, the $L_1$ and $L_\infty$ distances are equivalent.  However, for real-valued $f(x) \in [0,1]$, the triangle inequality over $\mathcal{P}(f)$ requires an $L_1$ bound.

Hence for our setting, $\eps$-uniform Glivenko-Cantelli convergence implies that a random sample $S \subset X$ of some size $k_\eps$ satisfies that $\sup_{p \in \R^d} |\kde_X(p) - \kde_S(p)| \leq \eps$; hence a random sample $S$ of size $k_\eps$ is an $\eps$-KDE coreset.  Thus our bound of $N_1(\eps,\mathcal{F}_K,X) = 2^{\tilde O(1/\eps^2)}$ implies that we can bound $k_\eps = \tilde O(1/\eps^4)$. 

It was already known~\cite{lopez2015towards,PT20} that $k_\eps = O((1/\eps^2)\log(1/\delta))$, with no dependence on $n$ or $d$ when the kernel $K$ is reproducing (e.g., for Gaussian or Laplace kernels).  So this reduction does not imply new $\eps$-KDE coreset results for this class of kernels.  
But our main result applies to ``simply computable'' kernels (defined below), includes the Epanechnikov, Triangle, Quartic, Triweight, and the truncated Gaussian, which are not reproducing.  

Moreover, by leveraging an intermediate result on semi-linked range spaces, we can also obtain bounds of size $k_\eps = \tilde O(1/\eps^6)$ (see Appendix \ref{app:KDE-samples}).

Next we discuss what the existing results~\cite{lopez2015towards,PT20} which show $k_\eps$ being independent of $n$ and $d$ imply about the size of the $\eps$-covering $N_\infty(\eps, \mathcal{F}_K, X)$.  Uniform Glivenko-Cantelli ensures that $ \lim_{m \to \infty} \frac{H_m(\eps,\mathcal{F}_K)}{m} = 0$.  This means  $H_m(\eps,\mathcal{F}_K) = \sup_{\substack{X \subset \mathcal{X} \\ |X|=m}} \log_2(N_1(\eps,\mathcal{F}_K,X)) = o(m)$, and so for $|X|=m$ then $N_1(\eps,\mathcal{F}_K,X) = 2^{o(m)}$.  Note that this bound can allow  $N_1(\eps,\mathcal{F}_K,X)$ to depend on the dimension $d$ as long as it is fixed.   Our result of $N_1(\eps,\mathcal{F}_K,X) = 2^{\tilde O(1/\eps^2)}$ is much stronger, as it implies no dependence on the $m = |X|$ or the dimension $d$.  

Finally, we back up to the notion of $V_\gamma$-shattering, and consider the implications of setting $\gamma = \eps$.  In our setting, a $V_\eps$-shattering dimension of $D$ would imply (using the contrapositive of the above definition) that for each $x \in X$ that for any set of $D+1$ (or more) points $A \subset X$, then no value $\alpha \in [0,1]$ can have all subsets $E \subseteq A$ $\eps$-separated.  That is, there must be some subset $E \subseteq A$ so there is not a function $f_E \in \mathcal{F}_K$ so for $x \in E$ that $f_E(x) \leq \alpha - \eps$, and for $x \in A \setminus E$ then $f_E(x) \geq \alpha + \eps$.  So a $V_\eps$-shattering of $\mathcal{F}_K$ dimension of $D = \tilde O(1/\eps^2)$ would imply that for $D+1$ or more points, that there would always be some subset $A$ that cannot be $\eps$-separated by a kernel $K(p,\cdot)$.  
This is not what our main technical lemma shows.  Instead, it uses a version of an $L_1$ distance, that shows that $K(p,\cdot)$ and $K(p',\cdot)$ differ \emph{on average} over $x \in X$ by at most $\eps$ for $p$ in an $\eps$-cover, and $p'$ as any point in $\R^d$.  Whereas the Alon \emph{et.al.} \cite{alon1997scale} paper uses a less discriminative $L_\infty$ distance that requires $K(p,\cdot)$ and $K(p',\cdot)$ to differ \emph{on all} points $x \in X$ by $\eps$.  

In summary, the famous Glivenko-Cantelli analysis of Alon \emph{et.al.} \cite{alon1997scale} provides a complete analysis of uniform convergence rates, but does not provide a finite sample size bound $k_\eps$ necessary to achieve an $\eps$ error.  In order to completely characterize this rate bound, they rely on a stronger $L_\infty$ type of distance between functional ranges; however, to provide a finite sample bound, we show we only need an $L_1$ variant of the distance between these functional ranges.  By restricting ourselves to this $L_1$ distance, we are able to show finite sample bounds for $k_\eps$ that are independent of $n$ and $d$ for a broad class of kernel range spaces defined over $n$ points in $\mathbb{R}^d$.

\section{Preliminaries:  Kernels, Range Spaces, and Covers}

There can be many definitions of a kernel, we start by specifying what properties are needed in this work.  A symmetric bi-variate function $K: \R^d \times \R^d \to \R$ is {\it centrally symmetric} if $K(p,x) = g(\|x-p\|)$, where $g:\R^{\geq 0} \to \R^{\geq 0}$ is a continuous function. We say that $K$ is $L$-Lipschitz if $g$ is.  Given $\eps>0$, the $\eps$-{\it critical radius} $r=r(\eps)$ of $K$ is the smallest positive real number such that $g(r') < \eps$ for any $r' > r$.  The ball $B_r(x) = \{p \in \R^d \mid \|p-x\| \leq r\}$  with $r=r(\eps)$ is then called the $\eps$-{\it critical ball} around $x$.  
We call $K$ a $(L,r)$-{\it standard kernel} if it is a non-negative, $L$-Lipschitz, centrally symmetric function with critical radius $r$ such that $r(\eps/2) = O(r(\eps))$ and for $\eps < 1/2$, $r(\eps) > C$ for some absolute constant $C>0$.  Standard kernels are continuous and bounded, which implies without loss of generality, by normalizing, we may assume that they take a maximum value of $1$.  The most common standard kernel is the Gaussian kernel $K(x,y) = e^{-\|x-y\|^2/\sigma^2}$; see others in Table \ref{table: L,r,k}, where the parameter $\sigma > 0$ is elsewhere assumed $\sigma=1$.  

A kernel $K$ is called {\it $k$-simply computable}, for some constant positive integer $k$, if for any $p,q,x \in \R^d$ and $\tau \in \R^+$, the inequality $|K(p,x) - K(q, x)| \geq \tau$ can be verified in $O(d^{k-1})$ steps using the ``simple'' arithmetic operations $+, -, \times$, and $/$, jumps conditioned on $>,\geq,<,\leq,=$, and $\neq$ comparisons, and $O(1)$ evaluations of the exponential function $z \mapsto e^z$ on reals. $K$ is called {\it simply computable} if it is $k$-simply computable for a constant $k$.  We will mostly work with simply computable kernels.  It is easy to see Gaussian, Epanechnikov, quartic and triweight kernels are $2$-simply computable;  Theorem \ref{thm:vc-dim: triangular} in Appendix \ref{app:VC-tri} shows triangle kernels are also $2$-simply computable.    It is not clear if Laplace kernels are simply computable, but they are positive definite, which we handle separately in Theorems \ref{thm:pd-kernels} and  \ref{thm: n,d-free-upper-bound-p-d}.  

\begin{table}[ht]
\small 
\centering
\begin{tabular}{llcccccc} 
\hline 
{\bf Kernel}    &               Rule: $K(x,y)$                  &       $L$ 	        &	   \hspace{-3mm}      $r$	            &	\hspace{-3mm} $k$ & $\eps$-cover sizes & \hspace{-3mm} Theorem \\ \hline 
Gaussian	    &   $e^{-\|x-y\|^2/\sigma^2}$             & $\frac{\sqrt{2/e}}{\sigma}$	& \hspace{-3mm} $\sigma \sqrt{\ln(1/\eps)}$  &	\hspace{-3mm} $2$	& $(\frac{1}{\eps})^{O(\frac{1}{\eps^2} \log^2(\frac{1}{\eps}))}$	& \hspace{-3mm} \ref{thm: n,d-free-upper-bound}, \ref{thm: n,d-free-upper-bound-p-d} \\ 
Laplace		    &   $e^{-\|x-y\|/\sigma}$           &   $1/\sigma$	        & \hspace{-3mm}	$\sigma \ln(1/\eps)$		&	\hspace{-3mm} 3  & $(\frac{1}{\eps})^{O(\frac{1}{\eps^2} \log^3(\frac{1}{\eps}))}$	& \hspace{-3mm} \ref{thm: n,d-free-upper-bound-p-d} \\ 
Epanechnikov	& $\max\{0, 1 - \frac{\|x-y\|^2}{\sigma^2}\}$     &   $2/\sigma$	        & \hspace{-3mm}	$\sigma\sqrt{1-\eps}$		&	\hspace{-3mm} $2$ & $(\frac{1}{\eps})^{O(\frac{1}{\eps^2} \log(\frac{1}{\eps}))}$ & \hspace{-3mm} \ref{thm: n,d-free-upper-bound} \\ 
Triangle		& $\max\{0, 1 - \frac{\|x-y\|}{\sigma}\}$         &   $1/\sigma$		    & \hspace{-3mm}	$\sigma(1-\eps)$		    &	\hspace{-3mm} $2$ & $(\frac{1}{\eps})^{O(\frac{1}{\eps^2} \log(\frac{1}{\eps}))}$	& \hspace{-3mm} \ref{thm: n,d-free-upper-bound} \\ 
Quartic     	& $\max\{0, 1 - \frac{\|x-y\|^2}{\sigma^2}\}^2$   &   $\frac{8}{3\sqrt{3}\sigma}$	    &	\hspace{-3mm} $\sigma \sqrt{1-\sqrt{\eps}}$	&	\hspace{-3mm} $2$ & $(\frac{1}{\eps})^{O(\frac{1}{\eps^2} \log(\frac{1}{\eps}))}$ & \hspace{-3mm} \ref{thm: n,d-free-upper-bound} \\ 
Triweight	    & $\max\{0, (1 - \frac{\|x-y\|^2}{\sigma^2})^3\}$ &   $\frac{96}{25\sqrt{5}\sigma}$	    & \hspace{-3mm}	$\sigma \sqrt{1-\sqrt[3]{\eps}}$	& \hspace{-3mm}	$2$ & $(\frac{1}{\eps})^{O(\frac{1}{\eps^2} \log(\frac{1}{\eps}))}$ & \hspace{-3mm} \ref{thm: n,d-free-upper-bound} \\ 
Trun-Gaussian  &   $\frac{1}{1-\tau}(e^{-\|x-y\|^2/\sigma^2} - \tau)$     & $\frac{\sqrt{2/e}}{\sigma}$	& \hspace{-3mm} $\sigma \sqrt{\ln(\frac{1}{\tau+(1-\tau)\eps})}$  & \hspace{-3mm}	$2$	& $(\frac{1}{\eps})^{O(\frac{1}{\eps^2} \log^2(\frac{1}{\eps}))}$ & \hspace{-3mm} \ref{thm: n,d-free-upper-bound} \\ \hline
\end{tabular}
\caption{Examples of standard kernels with $L$, $r$, $k$ values and their $d,n$-free $\eps$-cover sizes.}
\label{table: L,r,k}
\end{table}

\vspace{-5mm}

\subsection{Generalized Range Spaces and $\eps$-Covers}

Recall that the symmetric difference of two sets $A$ and $B$ is $A \symdiff B = (A \cup B) \setminus (A \cap B)$. Let $(X, \cA)$ be a range space, where $X\subset \mathbb{R}^d$ is a finite set, and let $\eps >0$. A range space $(X, \cA_{\symdiff})$ is called an $\eps$-{\it covering} of $(X, \cA)$ if $\cA_{\symdiff} \subset \cA$, and for any $A \in \cA$ there is an $A' \in \cA_{\symdiff}$ such that $|(X \cap A) \symdiff (X \cap A')| \leq \eps |X|$~\cite{PM2018}; that is the difference in elements that $A$ and $A'$ cover is $\leq \eps |X|$.

Consider the $n$-dimensional hypercube 
\[
\W^n = [0,1]^n = \{w = (w_1, \ldots, w_n) \in \mathbb{R}^n: 0 \leq w_i \leq 1 \ (1\leq i \leq n)\},
\]
equipped with the normalized $L_1$-distance  
\[
\textstyle d_{\symdiff}(w, w') = \frac{1}{n}\|w-w'\|_1 = \frac{1}{n} \sum_{i=1}^n |w_i - w_i'|,
\]
where $w = (w_1, \ldots, w_n), w' = (w_1', \ldots, w_n') \in \mathbb{W}^n$.

\subparagraph*{Generalized $\eps$-Covers.}
For a point set $X = \{x_1, \ldots, x_n\} \subset \R^d$ and a kernel $K: \R^d \times \R^d \to [0,1]$, a query point $p \in \R^d$ defines a signature vector
\[
R_p^{X, K} = (K(p, x_1), \ldots, K(p, x_n)) \in \W^n.  
\]
The {\it generalized symmetric difference distance} (with respect to $K$ and $X$) on $\R^d$ is defined by 
\[
d_{\symdiff}^{X, K}(p, q) = \textstyle d_{\symdiff}(R_p^{X, K}, R_q^{X, K}) = \frac{1}{|X|}\sum_{x \in X} |K(p, x) - K(q, x)|.
\]
Since the kernel $K$ will be fixed, for simplicity, we remove it from the superscripts of $R_p^{X, K}$ and $d_{\symdiff}^{X, K}(p, q)$.  We may also remove the superscript $X$ if there is no ambiguity.  For $S \subset X$ if we set $w_i=1$ when $x_i \in S$ and $0$ otherwise (i.e., $w$ is a corner of $\W^n$), and similarly for $S' \subset X$, then $d_{\symdiff}(w, w') = |S \symdiff S'|/n$.

For a kernel range space $(X,K)$, an \emph{$\eps$-cover} is a set $Q \subset \R^d$ such that for every $p \in \R^d$ there exists a $q \in Q$ such that $d_{\symdiff}^X(p, q) \leq \eps$.  If $K$ is $(L,r)$-standard, then 
\[
d_{\symdiff}^X(p, q) = \frac{1}{n}\sum_{i=1}^n |K(p, x_i) - K(q, x_i)| \leq \frac{L}{n} \sum_{i=1}^n \|p - q\| = L \|p - q\|;
\]
hence a sufficient condition for $\eps$-cover $Q$ is for all $p \in \R^d$ to have some $q \in Q$ so $\|p-q\| \leq \eps/L$. 

Next we can restrict this condition further to just query points $p$ in the critical ball defined by $K$ around each $x_i \in X$, i.e., $p \in \tilde{B} = \bigcup_{i=1}^n B_r(x_i)$, where $r$ is the $\eps$-critical radius of $K$. Otherwise, i.e. if $p, q \notin \tilde{B}$, both $K(p, x_i)$ and $K(q, x_i)$ are less than $\eps$ and so the inequality $d_{\symdiff}^X(p, q) \leq \eps$ holds trivially. In fact, one point $q \notin \tilde{B}$ can cover all points in $\R^d \setminus \tilde{B}$ as an $\eps$-cover. We denote this point by $q_{\infty}$. That is, if we get a $\tau$-cover for $X$ in the sense of metric spaces ($\tau$ depends on $\eps$ and $K$), then we have an $\eps$-cover for $(X, K)$.

\begin{theorem}\label{thm:Gaussian Kernel}
Consider a point set $X$ of size $n$ in $\R^d$ and a $(L, r)$-standard kernel $K(x,y)$. One can construct an $\eps$-cover of size $n \big(\frac{3Lr}{\eps}\big)^d$for the kernel range space $(X, K)$.

\end{theorem}
\begin{proof}
A metric space $\frac{\eps}{L}$-cover of $\tilde{B} = \cup_{i=1}^n B_r(x_i)$ will provide an $\eps$-cover of $(X, K)$.  The covering number, $N(V, \eps)$, of $V \subset \R^d$ is bounded by $\frac{\vol(V)}{\vol(B)} (\frac{3}{\eps})^d$, where $B$ shows the unit ball in $\R^d$.  The volume of a ball of radius $r$ in $\R^d$ is $\frac{\pi^{d/2}}{\Gamma(d/2+1)} r^d$, where $\Gamma$ is the Gamma function.  So each ball of radius $r$ can be covered by $\big(\frac{3r}{\eps/L}\big)^d$ points. Therefore, $n \big(\frac{3Lr}{\eps}\big)^d$ points are sufficient for an $\eps$-cover of $(X, K)$.
\end{proof}

\begin{corollary}\label{cor:Gau-Lap-Tri}
For point set $X \subset \R^d$ of size $n$, applying Theorem \ref{thm:Gaussian Kernel} and Table \ref{table: L,r,k}, these kernels admit $\eps$-covers for $(X, K)$: 

 \begin{tabular}{ll@{\hskip 10mm}ll}
      Gaussian & size: $n(3/\eps)^d \ln^{d/2}(1/\eps)$ & Laplace & size: $n(3/\eps)^d\ln^{d}(1/\eps)$ \\
      Triangle & size:  $n (3/\eps)^d$ & Epanechnikov & size: $n(3/\eps)^d$ \\
      Quartic & size: $n(3/\eps)^d$ & Triweight & size: $n(3/\eps)^d$ \\
 \end{tabular}
\end{corollary}

\section{$\eps$-Cover-Samples and Reducing the Number of Points}
We next build to the new definition of an $\eps$-cover-sample.  This is a subset of points that preserves the above gridding-based construction for an $\eps$-cover.  We start with the more well-known definition of an $\eps$-sample, and its existing generalization to kernels.  We observe this does not quite capture the right definition of a subset, so we evolve it to the $\eps$-cover-sample.  

An $\eps$-{\it sample} \cite{GeoAppAlg2011} for a range space $(X, \cA)$ is a set $S \subset X$ such that 
\[
\max_{A \in \cA} \bigg| \frac{|X \cap A|}{|X|} - \frac{|S \cap A|}{|S|} \bigg| \leq \eps.
\]
If $K \leq 1$ is a kernel defined on $X$, then, following \cite{joshi2011comparing}, an $\eps${\it-KDE-sample} (also called $\eps$-{\it sample for} $(X, K)$) is a set $S \subset X$ so 
\[
\max_{q \in \R^d} \bigg|\frac{1}{|X|} \sum_{x \in X} K(q, x) - \frac{1}{|S|}\sum_{s \in S} K(q, s) \bigg| \leq \eps.
\]

A kernel is said to be {\it linked}~\cite{joshi2011comparing} to a range space $(X, \cA)$ if for any possible input point $q \in \R^d$ and any value $v \in \R^+$ the super-level set of $K(\cdot,q)$ defined by $v$ is equal to some $H \in \cA$, i.e. $\{x \in \R^d: K(x, q) \geq v\} = H$. If $S$ is an $\eps$-sample for $(X,\cA)$, where $\cA$ is linked to $K$, then $S$ is also an $\eps$-KDE-sample~\cite[Theorem 5.1]{joshi2011comparing}.

An $\eps$-{\it cover-sample} for $X$ is a set $S \subset X$ such that for any $p, q \in \R^d$,
\[
    \big| d_{\symdiff}^X(p,q) - d_{\symdiff}^S(p,q) \big| \leq \eps.
\]
Appendix \ref{app:coreset} shows that an $\eps$-cover-sample is an $\eps$-KDE-sample, so 
an $\eps$-cover-sample is a generalization of an $\eps$-KDE-sample.  
However, we can only show an $\eps$-KDE-sample implies the following (note the absolute values outside the sum):
\[
\bigg|\bigg| \frac{1}{|X|} \sum_{x\in X} \big[ K(p,x) - K(q,x)\big]\bigg| - \bigg| \frac{1}{|S|} \sum_{s\in S}\big[K(p,s) - K(q,s)\big] \bigg| \bigg| \leq 2 \eps.
\]
Moreover, the example below shows that an $\eps$-cover-sample may not be an $\eps$-cover.

\begin{example}\label{exa2}
Fix $z \in \R$ and let $X=\{x_1, \ldots, x_n\} \subset \R$, where $x_1=\cdots = x_n = z$ (or small perturbations of $z$), and consider the Gaussian kernel $K(x,y)=e^{-\|x-y\|^2}$. Then $S=\{z\}$ will be an $\eps$-cover-sample for any $\eps >0$ since $K(p, x_i) = K(p, z)$ for all $i$ and all $p\in \R$. So, for any $p,q \in \R$, 
\[
\big|d_{\symdiff}^X(p, q) - d_{\symdiff}^S(p, q)\big| = \bigg|\frac{1}{n}\sum_{i=1}^n |K(p, x_i) - K(q, x_i)| - |K(p, z) - K(q, z)|\bigg| = 0 < \eps.
\]
Now let $p\in \R$ be arbitrary. If $S$ is an $\eps$-cover for $(X, K)$, then we should have $d_{\symdiff}^X(p, z) < \eps$ (remember that $S$ is a singleton and the only choice from $S$ is $z$) and so considering the fact that $R_z=(1, \ldots, 1)$ (since $x_1=\cdots = x_n = z$), we will need to have 
\[
|K(p, z) - 1| = \frac{1}{n}\sum_{i=1}^n |K(p, x_i) - 1| = \frac{1}{n}\sum_{i=1}^n |K(p, x_i) - K(z, x_i)| < \eps.
\]
However, this is impossible for an arbitrary $p \in \R$; and any $\eps$-cover $Q$ needs at least one point in each annulus $S_i = \{p: (i-1) \eps < K(p,z) \leq i \eps \}$ for $1 \leq i \leq 1/\eps$. 
\end{example}

\subsection{Semi-Linked Range Spaces}

The {\it semi-super-level set} of a kernel $K$ with respect to the points $p,q \in \R^d$ and $\tau \in \R^+$ is 
\[
R_{p,q,\tau} = \{x \in \R^d: |K(p, x) - K(q,x)| \geq \tau\}.
\]
Moreover, $K$ is said to be {\it semi-linked} to a range space $(\R^d, \cA)$ if $R_{p,q,\tau} \in \cA$ for any possible $p,q \in \R^d$, $\tau \in \R^+$.  We also say $\cA$ is semi-linked to $K$.  This is extending the idea of super-level sets and linking kernels to range spaces from \cite{joshi2011comparing}.  There are also $\eps$-KDE-samples of size $O(1/\eps^2)$ for characteristic kernels, either using a uniform random sample~\cite{lopez2015towards} or via an iterative greedy algorithm~\cite{bach2012equivalence,LLB2015}; see discussion in \cite{phillips2020near}.  
The size can be improved to $O(1/\eps)$ when $d$ is constant~\cite{tai2022optimal} for Gaussian kernels or for a bounded domain~\cite{karnin2019discrepancy}, or $O(\sqrt{d}/\eps \cdot \sqrt{\log(1/\eps)})$ for positive definite kernels~\cite{phillips2020near}.  

In the following theorem we extend the linking-based result to $\eps$-cover-samples that are now semi-linked to an appropriate range space.   The proof mostly follows the strategy of Joshi \emph{et al.} \cite{joshi2011comparing}, and is deferred to Appendix \ref{app:coreset-semi-linked}. The main idea is for any query based on values $p,q \in \R^d$, all points in $X$ can be sorted in descending order by their value in $|K(p,x)-K(q,x)|$.  For each subset in this order, it is guaranteed to have the appropriate error levels by the associated semi-linked range space.  Through some case analysis, as in \cite{joshi2011comparing}, one can show that the error cannot accumulate too much across different levels.

\begin{theorem}\label{thm:sym-coreset}
Let $S$ be an $\eps$-sample for $(X,\cA)$, where $\cA$ is semi-linked to a kernel $K$, where $K\leq 1$. Then $S$ is an $\eps$-cover-sample for $X$. 
\end{theorem}

Consider a range space $(X,\cA)$, and a kernel $K$ such that its critical radius is finite for any $\eps$ and $\cA$ is semi-linked to $K$. Then $\cA$ is linked to $K$, so this is a generalization of the linked range space result.

\begin{theorem}\label{thm:coreset}
Consider a $(L,r)$-standard kernel $K$ in $\R^d$. One can construct an $\eps$-cover with $s \big(\frac{Lr}{\eps}\big)^d$ points for the kernel range space $(X, K)$, where $s$ is the size of an $\eps/2$-sample for $(X,\cA)$, where $\cA$ is semi-linked to $K$. 
\end{theorem}
\begin{proof}
Let $S$ be an $\eps/2$-sample of size $s$ for $(X,\cA)$, where $\cA$ is semi-linked to $K$. Then 
\begin{equation} \label{eq7}
    \forall p, q\in \R^d, \ \big|d_{\symdiff}^X(p, q) - d_{\symdiff}^S(p, q)\big| \leq \eps/2.
\end{equation}
Applying Theorem \ref{thm:Gaussian Kernel} for $S$ we can get an $\eps/2$-cover $Q$ for $(S, K)$ of size $s\big(\frac{6Lr}{\eps}\big)^d$. Let $p \in \R^d$ be arbitrary. Choose $q \in Q$ such that $d_{\symdiff}^S(p, q) \leq \eps/2$.
Then utilizing \eqref{eq7} we get   
\[
d_{\symdiff}^X(p, q) \leq \big|d_{\symdiff}^X(p, q) - d_{\symdiff}^S(p, q)\big| + d_{\symdiff}^S(p, q) \leq \eps.
\qedhere \]
\end{proof}

\subsection{Bounding the VC-Dimension of Semi-Linked Range Spaces} \label{sec: vc-dim}

Given a range space $(X,\cA)$, a subset $Y \subset X$ is said to be {\it shattered} by $\cA$ if all subsets $Z\subset Y$ can be realized as $Z=Y \cap R$ for some $R \in \cA$. Then the {\it VC-dimension} of a range space $(X,\cA)$ is the cardinality of the largest subset $Y \subset X$ that can be shattered by $\cA$. 

Let $\cA_d = \{R_{p,q,\tau}: p,q\in \R^d, \tau > 0\}$, where $K$ is a standard kernel and $R_{p,q,\tau}$ is its semi-super-level set. Now consider the class of functions $H = \{h_a: \R^d \to \{0,1\} | a\in \R^N\}$, where $h_a(x) = h(a, x)$ and $h:\R^N \times \R^d \to \{0,1\}$ is a function.  This each $h_a( \cdot )$ defines a subset of $\R^d$ -- the points $x$ which evaluate to $1$.  Suppose $h$ is ``simply computable'', that is, computing $h(a,x)$ for any $a \in \R^N$ and $x\in \R^d$ requires no more than $t$ of the arithmetic operations, jumps conditions (described in Preliminaries) and comparisons, and requires $u$ times evaluation of the exponential function $z \mapsto e^z$.  Then Theorem 8.14 of \cite{AB2009} implies
\[
\dim((\R^d, H)) \leq d^2(u+1)^2 + 11d(u+1)(t+\log(9d(u+1))).
\]

In the appendix we have shown that $\dim((\R^d, \cA_1)) = 4$, $\dim((\R^d, \cA_2)) \geq 6$ and $\dim((\R^d, \cA_d)) \geq d+1$, where $K$ is any of the kernels listed in Table \ref{table: L,r,k}.  In order to get an upper bound on the VC-dimension of $\cA_d$ we employ the above simple operations theorem.

\begin{theorem} \label{vc-dim}
Let $\cA_d = \{R_{p,q,\tau}: p,q\in \R^d, \tau > 0\}$, where $K$ is a $k$-simply computable standard kernel and $R_{p,q,\tau} = \{x\in \R^d : |K(p,x) - K(q, x)| \geq \tau\}$. Then $\dim(\cA_d) = O(d^k)$.
\end{theorem}
\begin{proof}
For $p, q\in \R^d$, $\tau \in \R^+$ let $\chi_{p,q,\tau}^+$ and $\chi_{p,q,\tau}^-$ be the characteristic functions of the sets 
\[
R_{p,q,\tau}^+ = \{x\in \R^d : K(p,x) - K(q, x) \geq \tau\} \quad \text{and} \quad R^-_{p,q,\tau} = \{x\in \R^d : K(p,x) - K(q, x) \leq -\tau\},
\]
respectively. Then $h_{p,q,\tau} = \chi_{p,q,\tau}^+ + \chi_{p,q,\tau}^-$ is the characteristic function of $R_{p,q,\tau}$, i.e. $h_{p,q,\tau}(x) = 1$ if $x \in R_{p,q,\tau}$ and $0$ otherwise. Therefore, we have the class of functions 
\[
H=\{h_{p,q,\tau}: \R^d \to \{0,1\} | \, p,q \in \R^d, \, \tau \in \R^+\},
\]
(we set $N=2d+1$ here, for the coordinates to describe $p,q,\tau$). 

Because determining $|K(p,x) - K(q,x)| \geq \tau$ needs $O(d^{k-1})$ simple operations, one can easily observe that $h_{p,q,\tau}(x)$ can be computed by $t= O(d^{k-1})$ steps using above-mentioned simple operations and $u = O(1)$ evaluations of the  exponential function.  Therefore, by \cite[Theorem 8.14]{AB2009}, the VC-dimension of $H$ is upper bounded by $\dim(H) =O(d^k)$.  Hence, $\dim(\cA_d)=O(d^k)$. 
\end{proof}

For a range space $(X, \cA)$ with VC-dimension $\nu$, a random sample from $X$ 
of size $O((1/\eps^2)(\nu + \log 1/\delta))$ is an $\eps$-sample with probability at least $1 - \delta$~\cite{Tal1994, YPLS2001}. Using this fact and applying Theorem \ref{thm:coreset} we obtain the following corollaries.

\begin{corollary} \label{cor:coreset-eps-cover}
Let $K$ be a $k$-simply computable standard kernel on $\R^d$. A random sample from $X$ of size $O((1/\eps^2)(d^k + \log 1/\delta))$ is an $\eps$-cover-sample for $X$ with probability $\geq 1-\delta$.  
\end{corollary}

Now applying Theorem \ref{thm:coreset} gives the following result.

\begin{corollary}\label{cor:coreset1}
Let $K$ be a $k$-simply computable $(L,r)$-standard kernel on $\R^d$.  Then $O\big((d^k + \log(\frac{1}{\delta})) (6Lr)^d/\eps^{d+2} \big)$ points suffice to construct an  $\eps$-cover for $(X, K)$ with probability $\geq 1-\delta$. 
\end{corollary}

\section{An Upper Bound on $\eps$-Cover Size for High Dimensions}\label{sec:upper bound}

This section builds a dimension-free size upper bound for a kernel $\eps$-cover.  We use the following theorem, termed {\it $\eps$-terminal dimensionality reduction}  or {\it $\eps$-terminal JL}, from \cite{NN2019} (see \cite{ChJ2021} for an algorithmic version of terminal JL.  Appendix \ref{app:TJL} also gives the algorithm for computing terminal JL):

\begin{theorem} (\cite[Theorem 1.2]{NN2019}) \label{TJL}
Let $\eps \in (0, 1)$ and $X=\{x_1, \ldots, x_n\} \subset \R^d$ be arbitrary with $n>1$. Then there exists a function $f : \R^d \to \R^m$ with $m = O(\log(n)/\eps^2)$ such that for all $x_i \in X$ and \underline{all $p \in \R^d$}, $\|p-x_i\| \leq \|f(p) - f(x_i)\| \leq (1+ \eps) \|p-x_i\|$. 
\end{theorem}

\begin{lemma}\label{lem:pre-terminal-dim-red}
Let $\eps >0$, $K$ be a $(L,r')$-standard kernel, $r=r'(\eps/2)$, and $S\subset \R^d$ be a finite subset of $\R^d$. Let also $f: \R^d \to \R^m$ be the $\eps/(2 L r)$-terminal dimensionality reduction transform for $S$, where $m=O(L^2r^2\log(|S|)/\eps^2)$. Then for any $p \in \R^d$ the following holds: 
\[
\sum_{s \in S} |K(f(p), f(s)) - K(p, s)| < \frac{\eps}{2} |S|.
\]
\end{lemma}
\begin{proof}
Since $K$ is $L$-Lipschitz, using terminal dimensionality reduction property, for any $p\in \R^d$ and $s \in S$, we infer that 
\begin{equation}\label{min}
|K(f(p), f(s)) - K(p, s)| \leq L |\|f(p)- f(s) \| - \|p-s\|| \leq \frac{ \eps}{2r} \|p - s\|.
\end{equation}
Therefore, applying \eqref{min}, for any $p, q \in \R^d$ and $s \in S$ (note $|f(S)|=|S|$ as $f$ is invertible on $S$), we get 
\[
\begin{array}{ll} 
    \displaystyle \sum_{s \in S} |K(f(p), \!\!\!\!\!\! & f(s)) - K(p, s)| 
    \\ &
    \displaystyle = \sum_{\|p-s\| > r} |K(f(p), f(s)) - K(p, s)| + \sum_{\|p-s\| \leq r}  |K(f(p), f(s)) - K(p, s)|
    \vspace{1mm} \\ & 
    \displaystyle \leq \sum_{\|p-s\| > r}  \frac{\eps}{2} + \sum_{\|p-s\| \leq r} \frac{\eps}{2r} \|p-s\| \leq \sum_{s \in S} \frac{\eps}{2} = \frac{\eps}{2} |S|.
\end{array}
\]
\end{proof}

\begin{lemma}\label{lem:terminal-dim-red}
Let $\eps >0$, $K$ be a $(L,r')$-standard kernel, $r=r'(\eps/2)$, and $S\subset \R^d$ be a finite subset of $\R^d$. Let also $f: \R^d \to \R^m$ be the $\eps/(2 L r)$-terminal dimensionality reduction transform for $S$, where $m=O(L^2r^2\log(|S|)/\eps^2)$. Then for any $p, q \in \R^d$ the following holds: 
\[
|d_{\Delta}^{f(S)}(f(p), f(q)) - d_{\Delta}^S(p, q)| < \eps.
\]
\end{lemma}
\begin{proof}
Applying Lemma \ref{lem:pre-terminal-dim-red}, for any $p, q \in \R^d$ and any $s \in S$ (note $|f(S)|=|S|$ as $f$ is invertible on $X$), we get 
\[
\begin{array}{ll} 
    & d_{\Delta}^{f(S)} (f(p), f(q)) 
    \displaystyle = \frac{1}{|S|} \sum_{s \in S} |K(f(p), f(s)) - K(f(q), f(s))| \\ &
    \displaystyle \leq \frac{1}{|S|} \sum_{s \in S} \big(|K(f(p), f(s)) - K(p, s)| + |K(p, s) - K(q, s)| 
    \displaystyle 
    +|K(f(q), f(s)) - K(q, s)| \big)
     \\ &
    \displaystyle \leq \frac{1}{|S|} \Big[\frac{\eps}{2} |S| + \sum_{s \in S} |K(p, s) - K(q, s)| + \frac{\eps}{2} |S| \Big] 
    \displaystyle = d_{\Delta}^S(p, q) + \eps.
\end{array} 
\]
Similarly, we can show  
$d_{\Delta}^S(p, q) \leq d_{\Delta}^{f(S)}(f(p), f(q)) + \eps$. 
\end{proof}

The following corollary will help us in proving our main result (Theorem \ref{thm: n,d-free-upper-bound}), where we need to construct an $\eps$-cover for a set $S$ from an $\eps$-cover of its image under terminal dimensionality reduction transform.

\begin{corollary}\label{cor:terminal-dim-red}
Let $\eps >0$, $K$ be a $(L,r')$-standard kernel, $r=r'(\eps/16)$, $X=\{x_1, \ldots, x_n\} \subset \R^d$ and $S$ be an $\frac{\eps}{4}$-cover-sample for $X$. Let also $f: \R^d \to \R^m$ be the $\eps/(16Lr)$-terminal dimensionality reduction transform on $S$, where $m=O(L^2 r^2\log(|S|)/\eps^2)$. If $Q$ is an $\frac{\eps}{8}$-cover for $(f(S), K)$, then we can compute an $\eps$-cover for $(X, K)$ of size at most $|Q|$. 
\end{corollary}
\begin{proof}
Compute a naive $\frac{\eps}{8}$-cover $Q_S = \{p_1, \ldots, p_M\}$ for $S$ of size $M = |S| \big(\frac{24Lr'}{\eps}\big)^d$ by Theorem \ref{thm:Gaussian Kernel}.  
We say a point $p_i$ is covered by a point $q \in Q$ if $f(p_i) \in B_{\eps/8}(q)$, the $\eps/8$-radius ball in the metric space $(\W^{|f(S)|}, d_{\symdiff}^{f(S)})$.  
Process points $p_i \in Q_S$ one-by-one, putting some in a new set $Q'$.  To process a $p_i$, put it in $Q'$, find a point $q \in Q$ that covers $p_i$, and remove $q$ from $Q$.  Remove all $p_j \in Q_S$ from $Q_S$ that are also covered by $q$.  The process concludes when the whole $Q_S$ is processed.  We claim that $Q'$ is an $\eps$-cover for $(X, K)$; it is of size at most $|Q|$. 

Let $p\in \R^d$. Since $Q$ is an $\frac{\eps}{8}$-cover for $(f(S), K)$, there is $q \in Q$ such that $d_{\Delta}^{f(S)}(f(p), q) < \frac{\eps}{8}$.  Let $p_i \in Q_S$ be such that $d_{\Delta}^{S}(p, p_i) < \frac{\eps}{8}$. If $p_i \in Q'$, we are done.  Otherwise, there must be a $p_j \in Q_S$ with $j < i$ such that $f(p_i), f(p_j) \in B_{\eps/8}(q')$ for some $q' \in Q$, and $p_j$ is included in $Q'$ by the construction of $Q'$.  Thus $d_{\Delta}^{f(S)}(f(p_i), f(p_j)) < \frac{\eps}{4}$.  The proof will be complete if we show $d_{\Delta}^X(p, p_j) < \eps$.  Employing Lemma \ref{lem:terminal-dim-red} with $\eps/(16 L r)$ and $S$ we obtain $|d_{\Delta}^{f(S)}(f(p), f(p_i)) - d_{\Delta}^S(p, p_i)| \leq \frac{\eps}{8}$, and so $d_{\Delta}^{f(S)}(f(p), f(p_i)) < \frac{\eps}{4}$.  Therefore, 
\[
\textstyle d_{\Delta}^{f(S)}(f(p), f(p_j)) \leq d_{\Delta}^{f(S)}(f(p), f(p_i)) + d_{\Delta}^{f(S)}(f(p_i), f(p_j)) < \eps/2.
\]
Applying Lemma \ref{lem:terminal-dim-red} in a similar fashion we get $|d_{\Delta}^{f(S)}(f(p), f(p_j)) - d_{\Delta}^S(p, p_j)| \leq \frac{\eps}{8}$.  Combining last two inequalities we conclude that $d_{\Delta}^S(p, p_j) < \frac{5\eps}{8}$.  On the other hand, because $S$ is an $\frac{\eps}{4}$-cover-sample for $X$, we have $|d_{\Delta}^X(p, p_j) - d_{\Delta}^S(p, p_j)| \leq \frac{\eps}{4}$, which means that $d_{\Delta}^X(p, p_j) < \frac{7\eps}{8} < \eps$.
\end{proof}

\subsection{Input-Size and Dimension-Free Bound for $\eps$-Cover Size}

The bound we proved in Corollary \ref{cor:coreset1} on $\eps$-cover size is input size $(n = |X|)$-free but depends on the dimension $d$.  In Theorem \ref{thm:pd-kernels} we give an input-size- and dimension-free upper bound on the size of $\eps$-cover-sample for positive definite kernels using Rademacher complexity.  Then in Theorem \ref{thm: eps-cover-sample-free-bound} we obtain a similar bound for $k$-simply computable standard kernels.  These two theorems will result in input-size- and dimension-free upper bound for $\eps$-cover size.

\begin{theorem} \label{thm:pd-kernels}
Let $X = \{x_1, \ldots, x_n\} \subset \R^d$, $K$ be a positive definite bounded kernel on $\R^d$, 
and let $\delta \in (0,1)$ be the probability of failure. Then a random sample of size $m > \frac{1}{49 \eps^2} \log(\frac{1}{\delta})$ is an $\eps$-cover-sample for $X$ with probability $\geq 1 - \delta$. 
\end{theorem}
\begin{proof} 
For any $p, q \in \R^d$, clearly $\E_{x \sim X}[f_{p,q}(X)] = \frac{1}{n} \sum_{i=1}^n |K(p, x_i) - K(q,x_i)|$, where $\E$ denotes the expectation. Consider an i.i.d random sample $S = \{s_1, \ldots, s_m\}$ of $X$ of size $m$. 
Define the family of functions
\[
G = \{f_{p,q} : X \to [0, 1] \mid \, p, q \in \R^d, \, f_{p,q}(x) = |K(p, x) - K(q,x)| \ \text{for} \ x \in X\}.
\]
Applying the two-sided Rademacher complexity bound theorem (see Theorem 3.3 of \cite{MRT2018} for a one-sided bound) on $X$, $S$ and $G$ with probability of at least $1 - \delta$ we get
\begin{equation}\label{eq8}
\bigg|\frac{1}{n} \sum_{i=1}^n |K(p, x_i) - K(q,x_i)| -  \frac{1}{m} \sum_{j=1}^m |K(p, s_j) - K(q, s_j)| \bigg| \leq 2 \hat{\mathfrak R}_S(G) + 3 \sqrt{\frac{\log(4/\delta)}{2m}},
\end{equation}
where $\hat{\mathfrak R}_S(G)$ denotes the empirical Rademacher complexity of $G$ with respect to the sample $S$. 
On the other hand, by Theorem  \ref{thm:Rademacher-RKHS} we know that $\hat{\mathfrak R}_S(B_1^\cH(0)) \leq \frac{1}{\sqrt{m}}$, where $B_1^\cH(0)$ denotes the unit ball around the origin in RKHS.  Notice $K(p, \cdot)$ belongs to $B_1^\cH(0)$ for any $p$ in $\R^d$. Now, by inspection of the definition of the Rademacher complexity one can see that the Rademacher complexity of $B_1^\cH(0) - B_1^\cH(0)$ will be at most $\frac{2}{\sqrt{m}}$. Employing Talagrand’s contraction principle (see Lemma 5 in \cite{MZ2003}) for $B_1^\cH(0) - B_1^\cH(0)$ and absolute value function (which is $1$-Lipschitz), we observe that the Rademacher complexity of $G$ is at most $\frac{2}{\sqrt{m}}$.  Therefore, applying our notation in the paper, by \eqref{eq8} we obtain
\[
\textstyle |d_{\symdiff}^X(p, q) - d_{\symdiff}^S(p, q)| \leq 4/\sqrt{m} + 3 \sqrt{\log(4/\delta)/(2m)} \leq 7 \sqrt{\log(1/\delta)/m}.
\] 
Setting $7 \sqrt{\log(1/\delta)/m} < \eps$ gives $m > \frac{1}{49 \eps^2} \log(\frac{1}{\delta})$. 
\end{proof}

The similar bound for $\eps$-cover-sample size of $k$-simply computable standard kernels relies on the following observation, which is an easy application of triangle inequality. 

\begin{lemma}\label{lem:tri-eneq}
Let $S$ be an $\eps/2$-cover-sample of $X$ and $S'$ an $\eps/2$-cover-sample of $S$. Then $S'$ is an $\eps$-cover-sample of $X$.
\end{lemma}

The intuition behind the proof of the following theorem is recursively applying Lemma \ref{lem:tri-eneq}, by creating $\eps$-cover-samples of $\eps$-cover-samples, each of smaller size.  At the start of each step $i$ we have a size $n_i$ and dimension $d_i$.  We can apply terminal JL to reduce the dimension to $d_i' = O((1/\eps^2) \log n_i)$, and then Corollary \ref{cor:coreset-eps-cover} to create an $\eps$-cover-sample of size (roughly) $n_{i+1} = O((d_i'/\eps)^2) = O((1/\eps)^6 \log^2 n_i)$.  Combining these steps does not immediately remove the dependence on $n$ (or the initial $d$), but it does push the dependence on $n$ into the log term.  Applying this recursively the dependence on $n$ can eventually be eliminated, but at the cost of a $\log^*(n)$ error factor (since we accumulate $\eps$-error at each recursive step), which ultimately needs to be folded back into the size bound, adjusting $\eps' = \eps/\log^*(n)$.  
Instead we apply an inductive argument (inspired by the proof of Theorem 12.3 of \cite{Nabil2022}), so we only need to argue about one step.  We show that applying the reductions with sufficiently small error parameter $\eps$ it can be independent of $n$ and $d$.  It again uses Lemma \ref{lem:tri-eneq} but only once.  However, this argument is complicated by the two-stage approach because the dependence on $n$ and $d$ are linked, and reducing one relies on the other.  Like the recursive method sketched above, by combining them we can reduce the dependence on both terms.

\begin{theorem} \label{thm: eps-cover-sample-free-bound}
Let $\eps, \delta \in (0,1)$, consider a finite point set $X \subset \R^d$ and let $K$ be a $k$-simply computable $(L,r)$-standard kernel. Then with probability at least $1-\delta$, a random sample of size 
$
    O\big(\frac{1}{\eps^{2+2k}} L^{2k} r^{2k} \log^k(\frac{Lr}{\eps\delta})\big)
$
from $X$ is an $\eps$-cover-sample for $X$.  
\end{theorem}
\begin{proof}
Let $T(\eps, \delta, X)$ denote the least positive integer such that a uniform sample of $X$ of size $T(\eps, \delta, X)$ is an $\eps$-cover-sample of $X$ with probability at least $1-\delta$. We prove the theorem by induction on $\eps$. If $n \leq \frac{1}{\eps^{2+2k}}$, then $T(\eps, \delta, X) \leq n \leq \frac{1}{\eps^{2+2k}}$, and so $X$ would be an $\eps$-cover-sample of $X$ satisfying the claim of the theorem. Thus assume that $n > \frac{1}{\eps^{2+2k}}$. Let $S$ be an $\eps/2$-cover-sample of $X$ with probability at least $1-\delta/2$. Then employing Lemma \ref{lem:tri-eneq}, any $\eps/2$-cover-sample $S'$ of $S$, with probability at least $1-\delta/2$, would be an $\eps$-cover-sample of $X$ with probability $\geq 1-\delta$. This means that for any $\eps/2$-cover-sample $S'$ of $S$ we have 
\begin{equation}\label{eq:eps/2-cover-size}
    T(\eps, \delta, X) \leq T(\eps/2, \delta/2, S) \leq |S'|.   
\end{equation}
Let $f: \R^d \to \R^m$ be the $\eps'$-terminal dimensionality reduction transform for the set $S$ and $\eps' = \eps/(6Lr(\eps/6))$, where  $m=O(L^2 r^2 \log(|S|)/\eps^2)$. Now consider the range space $(f(S), \cA_m)$. By Theorem \ref{vc-dim}, $\dim(\cA_m)= O(m^k)$. Hence, a random sample $S''$ of size $n'' = (C_1/\eps^2)(m^k + \log 1/\delta)$ from $f(S)$ is an $\eps/6$-sample for $f(S)$ with probability at least $1 - \delta/2$ \cite{Tal1994, YPLS2001}, where $C_1$ is a sufficiently large constant. Now Theorem \ref{thm:sym-coreset} shows that $S''$ is an $\eps/6$-cover-sample for $f(S)$ with probability at least $1 - \delta/2$. However, since $f$ is invertible on $S$, we have $S'' = f(S')$, where $S' = \{s' \in S: f(s') = s'' \ \text{for some} \ s'' \in S''\}$. Let us show that $S'$ is an $\eps/2$-cover-sample for $S$. Let $p,q \in \R^d$ be arbitrary. Then with probability $\geq 1 - \delta/2$,
\[
\begin{array}{ll}
    |d_{\symdiff}^S\!\!\! & \!\!\!\! (p, q)  - d_{\symdiff}^{S'}(p, q)| \leq 
    |d_{\symdiff}^S(p, q) - d_{\symdiff}^{f(S)}(f(p), f(q))|  \vspace{1mm} \\ & 
    + |d_{\symdiff}^{f(S')}(f(p), f(q)) - d_{\symdiff}^{S'}(p, q)| + |d_{\symdiff}^{f(S)}(f(p), f(q)) - d_{\symdiff}^{f(S')}(f(p), f(q))| 
    \vspace{1mm} \\ &
    \leq \eps/6 + \eps/6 + \eps/6 = \eps/2,
\end{array}
\]
where we applied Lemma \ref{lem:terminal-dim-red} two times and utilized the fact that $f(S') = S''$ is an $\eps/6$-cover-sample for $f(S)$. Obviously, $|S'| = |S''|=n''$. By plugging in  $m=C_2 L^2 r^2 \log(|S|)/\eps^2$ for some constant $C_2>0$, we obtain
\[
|S'| = \frac{C_1}{\eps^2} \Big[\Big(\frac{C_2}{\eps^2}L^2r^2\log(|S|)\Big)^k + \log \frac{1}{\delta}\Big] \leq \frac{C_1 (C_2^k+1)}{\eps^{2+2k}} L^{2k}r^{2k} \log^k\big(\frac{|S|}{ \delta}\big).
\]
Since the above inequality holds for any $\eps/2$-cover-sample $S$ of $X$, we can infer that 
\[
|S'| \leq \frac{C_1 (C_2^k+1)}{\eps^{2+2k}} L^{2k} r^{2k} \log^k\big(\frac{T(\eps/2, \delta/2, X)}{\delta}\big).
\]
Therefore, applying inductive hypothesis and the fact that $k$ is constant and for any constant $a > 1$ and $x \geq a^{1/(a-1)}$, $\log^k(ax) \leq a^k \log^k(x)$, which can be easily observed, we get 
\[
\begin{array}{ll}
    \displaystyle \log^k\Big(\frac{\frac{2^{2+2k} C}{\eps^{2+2k}} L^{2k} r(\eps/2)^{2k} \log^k(\frac{4Lr(\eps/2)}{\eps \delta})}{\delta}\Big) & 
    \displaystyle \leq \log^k \bigg(\Big(\frac{2 C^{1/(2+2k)} L r(\eps/2) \log(\frac{4Lr(\eps/2)}{\eps \delta})}{\eps\delta}\Big)^{2+2k}\bigg) \vspace{1mm} \\ & 
    \displaystyle \leq (2+2k)^k C_3^k C^{k/(2+2k)} \log^k\Big(\frac{L r  \log(\frac{4Lr(\eps/2)}{\eps \delta})}{\eps \delta}\Big) \vspace{1mm} \\ & 
    \displaystyle \leq (2+2k)^k C_3^k \sqrt{C} \Big(\log\big(\frac{Lr}{\eps \delta}\big) + \log\log\big(\frac{C_3Lr}{\eps\delta}\big)\Big)^k \vspace{1mm}\\ & 
    \displaystyle \leq (2+2k)^k C_3^k \sqrt{C} \Big( 1.4 \log \big(\frac{C_3 Lr}{\eps\delta}\big) \Big)^k \vspace{1mm} \\ & 
    \displaystyle \leq 1.4^k (2+2k)^k C_3^{2k} \sqrt{C} \log^k \big(\frac{Lr}{\eps \delta}\big),
\end{array}
\]
and thus 
\[
\begin{array}{ll}
    |S'| \displaystyle \leq \frac{1.4^k (2+2k)^k C_1 (C_2^k+1) C_3^{2k} \sqrt{C}}{\eps^{2+2k}} L^{2k} r^{2k} \log^k \big(\frac{Lr}{\eps \delta}\big),
\end{array}
\]
where $C_3$ is a constant making sure that $4r(\eps/2) \leq C_3 r(\eps)$; recall that here $r = r(\eps)$ is the $\eps$-critical radius. 
Hence, if we put $C$ such that  $\sqrt{C} \geq 1.4^k (2+2k)^k C_1 (C_2^k+1) C_3^{2k}$, then by \eqref{eq:eps/2-cover-size} we conclude $T(\eps, \delta, X) \leq \frac{C}{\eps^{2+2k}} L^{2k} r^{2k} \log^k(\frac{Lr}{\eps \delta})$. 
\end{proof}

With the same proof technique as in Theorem \ref{thm: eps-cover-sample-free-bound} one can prove the following corollary, which gives an input-size- and dimension-free upper bound on the $\eps$-KDE-samples.  For characteristic kernels it is known that smaller such upper bounds of size $O(\frac{1}{\eps^2} \log \frac{1}{\delta})$ exist \cite{lopez2015towards,Phillips2013,CWS2010,bach2012equivalence,LLB2015}.  Our bound, however, applies for non-characteristic kernels as well. 
The only modification we need is applying super-level sets rather than semi-super-level sets.  Notice, in this setting, more kernels can be considered as $k$-simply computable such as Laplacian kernel, where  one can argue that it is $3$-simply computable.

\begin{corollary}\label{cor: kernel-eps-sample-size}
Let $\eps, \delta \in (0,1)$, consider a finite point set $X \subset \R^d$ and let $K$ be a $k$-simply computable $(L,r)$-standard kernel in the sense of super-level sets. Then with probability $\geq 1-\delta$, a random sample of size 
$O\big(\frac{1}{\eps^{2+2k}} L^{2k} r^{2k} \log^k(\frac{Lr}{\eps\delta})\big)$
from $X$ is an $\eps$-KDE-sample for $X$. 
\end{corollary}

Finally, we reach our goal for this section which was providing an input-size- and dimension-free upper bound on $\eps$-cover size. Recall that for most kernels the Lipschitz factor $L$ is a constant, and the critical radius $r$ is a constant or $\mathrm{polylog}(1/\eps)$.

\begin{theorem}\label{thm: n,d-free-upper-bound}
Let $\eps > 0$ and $X=\{x_1, \ldots, x_n\} \subset \R^d$, and $K$ be a $k$-simply computable $(L, r)$-standard kernel.  Then, with constant probability, we can compute an $\eps$-cover for $(X, K)$ of size $\big(\frac{Lr}{\eps}\big)^{O\big(\frac{L^2r^2}{\eps^2} \log(\frac{Lr}{\eps})\big)}$. 
\end{theorem}
\begin{proof}
Let $S$ be an $\eps/4$-cover-sample of $X$ of size $|S| = O(\frac{1}{\eps^{2+2k}} L^{2k} r^{2k} \log^k(\frac{Lr}{\eps}))$; which is guaranteed to exist by Theorem \ref{thm: eps-cover-sample-free-bound}, and fix the probability of failure $\delta = 0.1$, or any constant in $(0,1)$.  Let $f: \R^d \to \R^m$ be the $\eps'$-terminal dimensionality reduction transform for the set $S$ and $\eps' = \eps/(4Lr)$, where $m = O(L^2r^2 \log(|S|)/\eps^2) = O(L^2 r^2 \log({Lr}/{\eps})/\eps^2)$.

Then by Corollary \ref{cor:terminal-dim-red}, an $\eps/8$-cover $Q$ of $(f(S), K)$ will provide us with an $\eps$-cover of $(X, K)$ of size at most $|Q|$. Finally, by Theorem \ref{thm:Gaussian Kernel}, $f(S)$ admits an $\eps/8$-cover of size 
$O((\frac{24}{\eps})^{m+2+2k} L^{m+2k} r^{m+2k} \log^k(\frac{Lr}{\eps})) = (\frac{Lr}{\eps})^{O(L^2r^2 \log(\frac{Lr}{\eps})/\eps^2)}$ for $(X, K)$. 
\end{proof}

By the discussion in Section \ref{sec:Glivenko-Cantelli} we conclude the following corollary, which interestingly is a $k$-free improvement upon Corollary \ref{cor: kernel-eps-sample-size}.

\begin{corollary}\label{cor: k-free-kernel-eps-sample-size}
Let $\eps, \delta \in (0,1)$, consider a finite point set $X \subset \R^d$ and let $K$ be a $k$-simply computable $(L,r)$-standard kernel in the sense of super-level sets.  Then with probability $\geq 1-\delta$, a random sample of size 
$\tilde{O}\big(\frac{1}{\eps^{4}}\big)$ from $X$ is an $\eps$-KDE-sample for $X$. 
\end{corollary}

We include a similar upper bound for positive definite kernels too.

\begin{theorem}\label{thm: n,d-free-upper-bound-p-d}
Let $\eps > 0$ and $X=\{x_1, \ldots, x_n\} \subset \R^d$, and $K$ be a positive definite $L$-Lipschitz kernel with critical radius $r$.  Then, with constant probability, we can compute an $\eps$-cover for $(X, K)$ of size $\big(\frac{Lr}{\eps}\big)^{O(L^2r^2 \log(\frac{1}{\eps})/\eps^2)}$. 
\end{theorem}
\begin{proof}
Let $S$ be an $\eps/4$-cover-sample of $X$ of size $|S| = O(\frac{1}{\eps^{2}})$; which is guaranteed to exist by Theorem \ref{thm:pd-kernels} assuming constant probability of failure $\delta$.  Let $f: \R^d \to \R^m$ be the $\eps'$-terminal dimensionality reduction transform for the set $S$ and $\eps' = \eps/(4Lr)$, where $m = O(L^2r^2 \log(|S|)/\eps^2) = O(L^2 r^2 \log({1}/{\eps})/\eps^2)$. 
Then by Corollary \ref{cor:terminal-dim-red}, an $\eps/8$-cover $Q$ of $(f(S), K)$ will provide us with an $\eps$-cover of $(X, K)$ of size at most $|Q|$. Finally, by Theorem \ref{thm:Gaussian Kernel}, $f(S)$ admits an $\eps/8$-cover of size $O((\frac{24}{\eps})^{m+2} L^{m} r^{m}) = (Lr / \eps)^{O(L^2r^2 \log(\frac{1}{\eps})/\eps^2)}$ for $(X, K)$. 
\end{proof}

Considering the fact that the Gaussian kernel is $(1, \sqrt{\ln(1/\eps)})$-standard and positive definite, triangle kernel is $(1, 1)$-standard, Epanechnikov, triangle, quartic and triweight kernels are $(2, 1)$-standard, and the Laplace kernel is $1$-Lipschitz and positive definite with critical radius $\ln(1/\eps)$, the following corollary is an immediate consequence of Theorems \ref{thm: n,d-free-upper-bound} and \ref{thm: n,d-free-upper-bound-p-d}.  Notice we assumed $\sigma = 1$.

\begin{corollary}\label{cor: n,d-free-upper-bound-Gau}
Let $\eps>0$ and $X \subset \R^d$ be of size $n$. There exist a set of size $(\frac{1}{\eps})^{O(\frac{1}{\eps^2} \log^2(\frac{1}{\eps}))}$ for Gaussian/truncated Gaussian kernel, a set of size $(\frac{1}{\eps})^{O(\frac{1}{\eps^2} \log^3(\frac{1}{\eps}))}$ for Laplace kernel, and a set of size $(\frac{1}{\eps})^{O(\frac{1}{\eps^2} \log(\frac{1}{\eps}))}$ for Epanechnikov, triangular, quartic and triweight kernels that are $\eps$-covers of $(X, K)$.
\end{corollary}

\subparagraph*{Algorithm and run time.}
The result in Theorem \ref{thm: n,d-free-upper-bound} is constructive with a runtime described in the next theorem. 
Recall, despite the involved analysis, the algorithm to find the $\eps$-cover $Q'$ of $X$ is quite simple:  
(1) create a random sample $S \sim X$; 
(2) create a terminal JL map $f : \R^d \to \R^m$ for $S$; 
(3) map $S' \leftarrow f(S)$;
(4) apply Theorem \ref{thm:Gaussian Kernel} on $S'$ in $\R^m$ to get $Q \subset \R^m$.  
(5) find $Q' \subset \R^d$ from $Q$.  
Analyzing the runtime of most steps is straightforward.  The most intricate part is step (5) since $f$ is only invertible on $S$.  This is handled via the procedure described in the proof of Corollary \ref{cor:terminal-dim-red}, whereby we create a naive $\eps$-cover $Q_S \subset \R^d$ (via Theorem \ref{thm:Gaussian Kernel}), and make sure to only place one point $p_i$ from $Q_S$ into the final $\eps$-cover $Q' \subset \R^d$ for each $q \in Q \subset \R^m$.

\begin{theorem}\label{thm:runtime}
For a size $n$ point set $X \subset \R^d$ and $k$-simply computable, $(L, r)$-standard kernel $K$ we can compute an $\eps$-cover for $(X,K)$ of size $N = (Lr/\eps)^{ O(L^2 r^2 \log(L r/\eps)/\eps^2)}$ in time $(Lr/\eps)^{d + O(\frac{1}{\eps^2} \log \frac{Lr}{\eps})}$, where we assume $k$ is a constant.
\end{theorem}
\begin{proof}
Assuming we can draw a random sample in $O(1)$ time, step (1) takes $O(|S|) = O(d^k/\eps^2)$ time. 
Creating a terminal JL map $f : \R^d \to \R^m$ for $S$ in steps (2) and (3) needs $O(d |S| \log(|S|)/\eps^2)$ time (see Appendix \ref{app:TJL}).  
In step (4) applying Theorem \ref{thm:Gaussian Kernel} on $f(S)$ requires $O(|S| (Lr/\eps)^m)$ time, where $m=O(\log(|S|)/\eps^2)$.  Similarly, in step (5) creating an $\eps$-cover $Q_S \subset \R^d$ for $S$ (in the proof of Corollary \ref{cor:terminal-dim-red}) requires $O(|S| (3Lr/\eps)^d)$ time.   
Then we need to calculate $f(p_i)$ for $p_i \in Q_S$ in Corollary \ref{cor:terminal-dim-red}, which needs $O(|Q_S| \sqrt{d}(|S|^3+d^3)\log(1/\eps))$ time (see Appendix \ref{app:TJL}).  
In addition, computing each distance $\|f(p_i) - q\|$ for any $p_i \in Q_S$ and $q \in Q$ needs $O(m)$ time.  Thus, noting that the process stops when $Q$ is empty, these distance calculations need $O(m|Q||Q_S|)$ time.  Therefore, putting all together, the run time of obtaining an $\eps$-cover of size $N$ needs $O(d |S| \log(|S|)/\eps^2 + |Q_S| \sqrt{d}(
|S|^3+d^3)\log(1/\eps) + |S|^2 + m |Q||Q_S|)$ time.  Substituting $|S| = O\big(\frac{1}{\eps^{2+2k}} L^{2k} r^{2k} \log^k(\frac{Lr}{\eps})\big)$ (by Theorem \ref{thm: eps-cover-sample-free-bound} assuming a constant probability),  
$|Q| = |S| (3Lr/\eps)^m = |S| (Lr/\eps)^m$, $|Q_S| = |S| (3Lr/\eps)^d$ and by simplifying we end up with $\left(Lr/\eps\right)^{d+ O(\log(Lr/\eps)/\eps^2)}$ time. 
\end{proof}

We leave as an open question if we can remove the $1/\eps^d$, so the runtime is $(\frac{1}{\eps})^{\mathrm{poly} (\frac{1}{\eps})}$.  This would seem to require a way to invert terminal JL for any point in $\R^m$.

\section{A Lower Bound on $\eps$-Cover Size for Gaussian and Laplace Kernel}
\label{sec:LB}

We now provide a lower bound, nearly matching the upper bound on $\eps$-cover size shown in Corollary \ref{cor: n,d-free-upper-bound-Gau} for the Gaussian kernel; it also applies to the Laplace kernel. First we define criteria on a set of $d$ spheres in $\R^d$ that can generate exactly two points in their intersections.  Then we use this to design a point set that provides the desired lower bound.

\begin{figure}
    \centering
    \includegraphics[width=\linewidth]{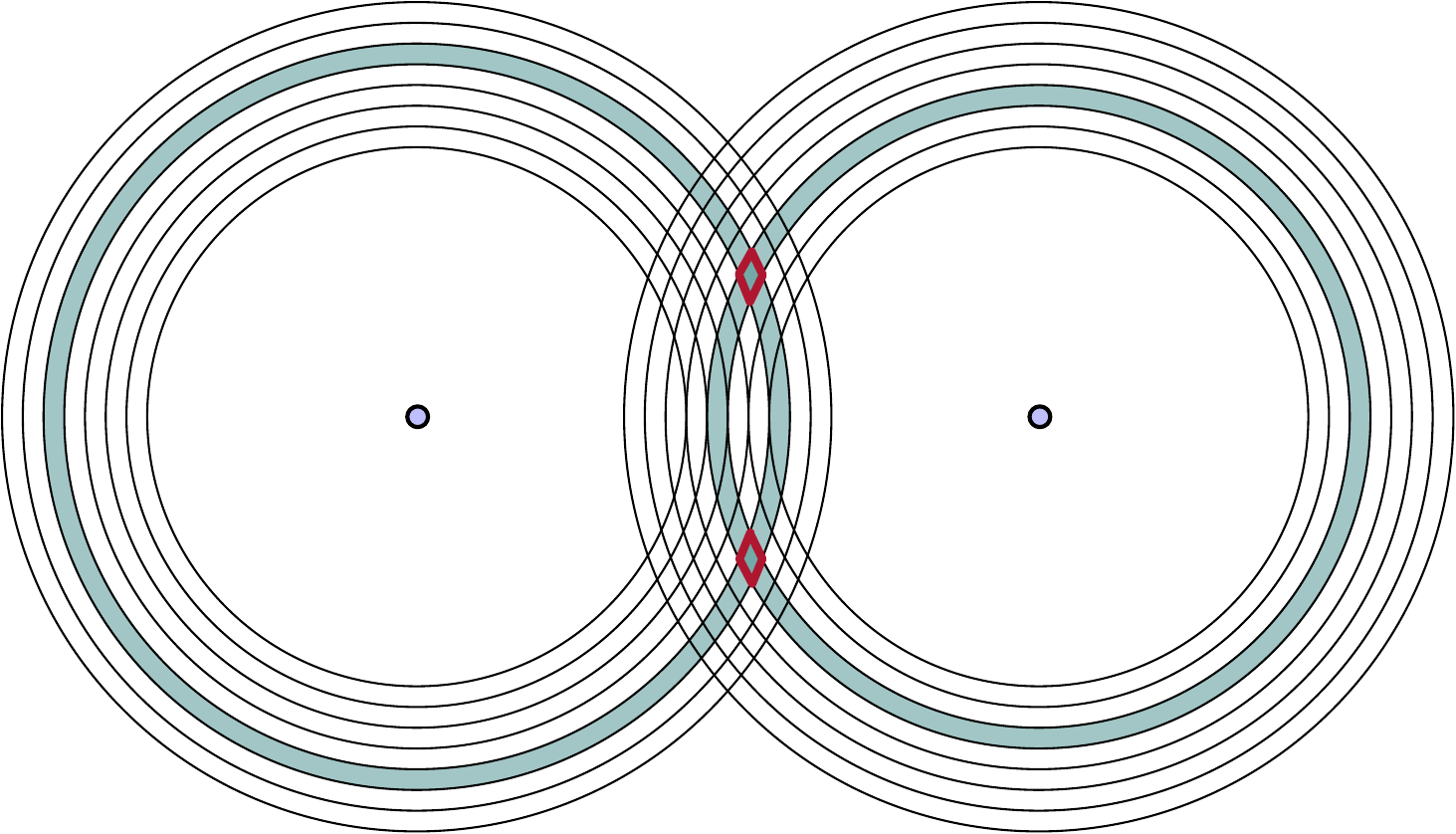}
    \caption{Illustration of intersection of annuluses around two blue points.  The two green annuluses intersect forming 2 grid-like cells in red.}
    \label{fig:annuluses}
\end{figure}

The following lemma, as we will see in Lemma \ref{lem: sphere intersection count}, will help us to make sure that every $d$-sphere in a collection of $1/\eps$ same-centered spheres will intersect $1/\eps^{d-1}$ spheres from another collection of spheres, each in 2 points. This will inductively provide us with $2/\eps^d$ grid-like cells, where each cell is obtained by intersecting $d$ annuluses formed by pairs of same-centered but different radius spheres with other annuluses. For instance, in $\R^2$ each grid-like cell is created as an intersection of 2 strips formed by pairs of same-centered circles.  This is illustrated in Figure \ref{fig:annuluses}. So, the lemma makes it easy to count the number of grids generated in this way as we need to consider them for the lower bound on $\eps$-cover.  The proof is deferred to the appendix (see Lemma \ref{lem: sphere intersection-proof}).

\begin{lemma} \label{lem: sphere intersection}
Let $S_1, \ldots, S_d$ be $d$-spheres in $\R^d$, where $S_k$ is centered at $e_k$ with radius $R_k \geq 1$ ($e_k$ is the standard $k$-th basis vector in $\R^d$). Assume that there are $R \geq 1$ and $\delta \in (0,1/d)$ such that $|R_k^2 - R^2| < \delta$ for $k=1, \ldots, d$. Then $|\bigcap_{k=1}^d S_k| = 2$.
\end{lemma}

Crucially, the restriction that $\delta < 1/d$ in Lemma \ref{lem: sphere intersection} will lead the next lemma to only obtain a $(1/\eps)^{\Omega((1-\lambda)d)}$ size bound when $d = O(1/\eps^\lambda)$ for any $\lambda \in (0,1)$.

\begin{lemma} \label{lem: sphere intersection count}
Let $\eps \in (0, 1/3)$ and $d < \frac{1}{e^{3-\lambda}} \frac{1}{\eps^\lambda}-\frac{1}{e}$ for some $\lambda \in (0,1)$. Then there are $\Omega(1/\eps^d)$ $d$-way intersections among  $d$-spheres $S_k$ in $\R^d$, where $S_k$ is centered at $e_k$ with radius $R_k = \ln(\frac{1}{i\eps})$ for some $k=1, \ldots, d$ and any integer $i$ in $[\frac{1}{(e + 1/d) \eps}, \frac{1}{e \eps}]$. Hence, there are at least  $(\frac{2}{\eps})^{(1-\lambda)d}$ grid-like cells obtained through the intersection of annuluses obtained by these spheres. Moreover, if $d$ is constant, then the lower bound can be improved to be $\Omega(1/\eps^d)$. 
\end{lemma} 
\begin{proof}
We will show that $d$-spheres in this setting satisfy the assumptions of Lemma \ref{lem: sphere intersection}.  
For any integer $i \in \big[\frac{1}{(e + 1/d) \eps}, \frac{1}{e \eps} \big]$, using $R_i^2=\ln (\frac{1}{i\eps})$ and $R=1$, we can infer 
\[
    |R^2 - R_i^2| 
    = \Big|1 - \ln \frac{1}{i \eps}\Big| 
    \leq \Big| 1 - \ln\big(e + \frac{1}{d}\big)\Big|
    < \frac{1}{d}, 
\]
where the last step follows by $x < e^x$.  Therefore, by Lemma \ref{lem: sphere intersection}, each sphere with radii $R_i$ centered at $e_k$ will intersect any sphere with radii $R_j$ centered at $e_\ell$, where $i, j$ are integers in $[\frac{1}{(e + 1/d) \eps}, \frac{1}{e\eps}]$ and $k \neq \ell$. Note 
\[
    \frac{1}{e\eps} - \frac{1}{(e + \frac{1}{d})\eps} = \frac{ 1}{e(ed + 1)\eps} > \frac{e^{1-\lambda}\eps^{\lambda}}{\eps} = \Big(\frac{e}{\eps}\Big)^{1 - \lambda}.
\]
It means that there would be at least $(\frac{e}{\eps}-1)^{(1-\lambda)d}$ $d$-way intersections.  Consequently, there would be  at least $(\frac{e}{\eps}-2)^{(1-\lambda)d} > (\frac{2}{\eps})^{(1-\lambda)d}$  
grid-like cells obtained through intersection of annuluses formed by consecutive spheres on each axis with annuluses centered on other axes, as illustrated in Figure \ref{fig:annuluses}.

If $d$ is fixed, then the length of the interval $[\frac{1}{(e + \frac{1}{d})\eps}, \frac{1}{e\eps}]$ would be  a constant fraction of $1/\eps$, leading to the lower bound of $\Omega(1/\eps^d)$.
\end{proof}

We remark that in the proof of Lemma \ref{lem: sphere intersection count} if we use $R_i = \ln(\frac{1}{i\eps})$, the lemma holds true.

The following theorem gives a lower bound on the kernel $\eps$-cover size.

\begin{theorem} \label{thm: eps-cover count}
Let $\eps \in (0, 1/3)$ and $d < \frac{1}{e^{3-\lambda}} \frac{1}{\eps^\lambda}-\frac{1}{e}$ for some constant $\lambda \in (0,1)$ and let $X=\{e_1, \ldots, e_d\} \subset \R^{d}$ be the vertices of the standard (d-1)-simplex (i.e. $e_k$ is the $k$-th basis vector). Let also  $K(x,y)=e^{-\|x-y\|^2}$ (or $K(x,y)=e^{-\|x-y\|}$). 
Then the size of any $\eps$-cover for $(X, K)$ is at least $(1/\eps)^{\Omega(1/\eps^\lambda)}$. 
If $d$ is a constant, then the size of any $\eps$-cover for $(X, K)$ is at least $\Omega(1/\eps^d)$.
\end{theorem}
\begin{proof}

We need to start by constructing objects on the grid-like structure we call super-cells. 
Let the $d$-way intersections of the spheres be the nodes in a graph.  Two nodes are connected by an edge if they share $d-1$ spheres and for the last dimension of the nodes, the associated spheres are consecutive in the radius ordering.  Then given a node $p$ (which corresponds with a point in $\R^d$), we can define an $L_1$-distance to other nodes in the graph as the minimum number of edges one needs to traverse to get from $p$ to another node $q$.  A super-cell $S_p$ contains all nodes within an $L_1$-distance of $\eps d$.  However, the super cell $S_p$ contains not just nodes, but also the part of $\R^d$ that one can reach from any node in a super-cell without crossing a sphere boundary.   
So one can think of a super-cell as a collection of cells from the grid-like structure that are reachable by moving into one of the $2^d$ cells incident to $p$, and then including face-incident cells to those within $\eps d$ additional steps.  

Next we want to argue that if there are no points $q$ in a set $Q$ that intersect a super-cell $S_p$, then for all $q \in Q$ that $d_{\symdiff}^X(p,q) > \eps$.  Thus, if $Q$ is an $\eps$-cover it must hit (there must exist some $q \in Q$ so $q \in S_p$) the super-cell $S_p$.  To see this, consider any point $q \in S_p$.  One can reach $q$ in a sequence of moves from $p=p_0$ to $p_1$ and recursively from $p_j$ to $p_{j+1}$.  The first $j_*$ steps for $j_* \leq \eps d$ moves must be along the edges of the grid-graph.  These steps have the property that for some dimension $k$ we change $|K(e_k,p_j) - K(e_k,p_{j+1})| = \eps$, and for every other dimension $k'$ we have $|K(e_{k'},p_j) - K(e_{k'},p_{j+1})| = 0$.  The last step from $p_{j_*}$ to $q$ must have $|K(e_k,p_{j_*}) - K(e_k, q)| \leq \eps$ for all dimensions $k$.  Now $d_{\symdiff}^X(p,q) = \frac{1}{d} \|R_p^X - R_q^X\|_1$ is $\frac{1}{d}$ times the sum of all changes in $|(R_p^X)_k - (R_q^X)_k| = |K(e_k,p) - K(e_k,q)|$.  The first $j_* \leq \eps d$ steps of the path captures any one-dimensional change in increments of $\eps$, and the last step all residual changes in any coordinate less than $\eps$.  If the sum of these changes is less than $\eps d$, then it must be captured by some path. Therefore, a point $q$ belongs to $S_p$ if and only if it can be captured by some path starting at $p$ with the sum of above-mentioned changes less than $\eps d$.  The latter is equivalent to $d_{\symdiff}^X(p,q) \leq \eps$.

Now we need to provide an upper-bound of the size of a super-cell in terms of cells in the grid-like structure.  Then we can lower bound the size of the $\eps$-cover by the number of cells of the grid-like structure divided by the size of a super-cell.  
To do so, we divide a super-cell into $2^d$ orthants from $p$, so each path in some orthant only allows each dimension $k$ to either increment or decrement.  
In addition, for each of the $2^d$ orthant choices, this determines which cell $q$ moves into in the last step where it deviates from the grid-graph:  it moves into the same-oriented incident orthant from $p_{j_*}$.  
By vector addition commutativity, we can take this step first to move from $p$ to one of the incident $2^d$ cells, and then the remaining steps move to face-incident cells in the grid-like structure.  
The number of steps $j_*$ can be between $0$ and $\eps d$.  Each step has $d$ choices.  So after fixing one of the $2^d$ orthants, the total number of distinct paths is $\sum_{j=0}^{\eps d} d^j \leq d^{\eps d+1}$ for $d > 1$.  Note this is an overcount since two paths may end up in the same location by changing the order of steps.  Regardless, we will use $2^d \cdot d^{\eps d+1}$ as an upper bound on the size of a super-cell.

Finally, by a volume argument we can lower bound the size of an $\eps$-cover as the number of cells in a grid-like structure, which is at least $(2/\eps)^{(1-\lambda)d}$ by Lemma \ref{lem: sphere intersection count}, divided by the number of cells in a super-cell which is at most $2^d \cdot d^{\eps d+1}$.  Further, by Lemma \ref{lem: sphere intersection count} we can use dimension as large as $d = \frac{1}{e^{3-\lambda}} \frac{1}{\eps^\lambda} - \frac{1}{e}-1$.  We have
\begin{align*}
\frac{(2/\eps)^{(1-\lambda)d}}{2^d \cdot d^{\eps d+1}}
&=
\frac{(2/\eps)^{(1-\lambda)(\frac{1}{e^{3-\lambda}} 1/\eps^{\lambda} - \frac{1}{e}-1)}}{2^{\frac{1}{e^{3-\lambda}} 1/\eps^\lambda - \frac{1}{e}-1} \cdot (\frac{1}{e^{3-\lambda}} 1/\eps^\lambda - \frac{1}{e}-1)^{\eps(\frac{1}{e^{3-\lambda}} 1/\eps^\lambda - \frac{1}{e}-1)+1}}
\\ & \geq
\frac{(2/\eps)^{(1-\lambda)(\frac{1}{e^{3-\lambda}} 1/\eps^{\lambda} - \frac{1}{e}-1)}}{2^{1/\eps^\lambda} \cdot (1/\eps^\lambda)^{\eps(1/\eps^\lambda)+1}}
\\ & =
2^{(1-\lambda)(\frac{1}{e^{3-\lambda}} 1/\eps^{\lambda} - \frac{1}{e}-1) - 1/\eps^\lambda}        \cdot
(1/\eps)^{(1-\lambda)(\frac{1}{e^{3-\lambda}} 1/\eps^{\lambda} - \frac{1}{e}-1) - \lambda \eps^{1-\lambda} + \lambda}
\\ & = 2^{\eps^{-\lambda}((1-\lambda)e^{\lambda-3}-1) - (1-\lambda)\frac{1}{e} - (1-\lambda)} 
      \cdot 
     (1/\eps)^{\eps^{-\lambda}((1-\lambda)e^{\lambda-3}) - \lambda \eps^{1-\lambda} + 2\lambda - (1-\lambda)\frac{1}{e} - 1}
\\ & = (1/\eps)^{\Omega(1/\eps^\lambda)}.
\end{align*}
The bound for constant $d$ can be obtained similarly. 
\end{proof}

Finally, notice that assuming a constant dimension $d$, the upper bound of $O(\log^{d/2}(1/\eps)/\eps^d)$ in Theorem \ref{thm:Gaussian Kernel} is up to logarithmic factors tight with respect to the lower bound of $\Omega(1/\eps^d)$.

\section{A Lower Bound on the $\eps$-Cover for Combinatorial Range Spaces} \label{app:lower-bound-eps-cover}

The lower bound given by Haussler for $\eps$-cover size is $\big(\frac{n}{2e(k+d)}\big)^{d}$, which is designed for a special range space $(X, \cR)$ with VC-dimension $d$, where $n = s d$ for some integer $s$ and $1 \leq k \leq n$, where $\eps = k/n$.  But it does not apply specifically to any common geometric range spaces like those defined for points in $\R^d$ and by half-spaces, balls, or fixed-radius balls.  

We address this for large $d$ case for half-spaces by providing a new $\eps$-cover size lower bound that is roughly $1/\eps^d$, and thus cannot be similar to what we obtained for kernel range spaces.  We then, via a discussion in Appendix \ref{app:H=frB=B}, argue this lower bound also holds for ball and fixed-radius ball range spaces.

\begin{theorem} \label{thm:lower-comb-eps-cover}
Let $\eps \in (0, 0.3)$, $n = d$ and $X = \{e_1, \ldots, e_d\}$ be the vertices of the standard $(d-1)$-simplex in $\R^d$.  Then one needs at least $M$ points as an $\eps $-cover for $(X, \cH)$, where $\cH$ denotes the half-space ranges and 
\begin{itemize}
    \item[(i)] $M = 2^d$ if $d \leq 1/\eps$,
    \item[(ii)] $M = 2^{(1 - \eps \log_2(e/\eps))d}$ if $d > 1/\eps$.  Notice, in this case, $1 - \eps \log_2(e/\eps) \to 1$ and so $M\to 2^d$ as $\eps \to 0$. 
\end{itemize}
\end{theorem}
\begin{proof}
$(i)$ If $d \leq 1/\eps$, in order to get an $\eps$-cover, one needs to consider all $2^d$ subsets of $X$, since any two half-spaces $h$ and $h'$ which contain a different subset of points, say $A$ and $B$, have $d_{\symdiff}^X(h, h') = \frac{1}{n} |A \symdiff B| > \eps$.  Therefore, in this case there is only one $\eps$-cover $Q = 2^X$. 

$(ii)$ Consider a half-space $h$ and a length $d$ binary vector $a = (a_1, \ldots, a_d)$ associated to $h$ by means of $a_i = 1$ if $x_i \in h$ and $a_i = 0$ if $x_i \notin h$.  We need to count the number of ranges that differ in at least $\eps d + 1$ points.  It means that we are allowed to flip $a_i$'s up to $\eps d$ times and this flipping does not affect our counting process.  Thus there are at most  
\[
N = {d \choose 0} + {d  \choose 1} + \cdots + {d  \choose \eps d}
\]
other ranges within an $\eps n = \eps d$ distance of $h$.  By a counting argument, any $\eps$-cover $Q$ will need at least $2^d/N$ elements.  We can upper bound $N$ by the well-known inequality $N \leq (\frac{ed}{\eps d})^{\eps d} = (\frac{e}{\eps})^{\eps d}$. Hence, 
\[
|Q| \geq \frac{2^d}{N} \geq \frac{2^d}{(e/\eps)^{\eps d}} = 2^{(1 - \eps \log_2(e/\eps))d}. \hfill \qedhere
\]
\end{proof}

\begin{corollary}\label{cor:lower-comb-eps-cover}
When $n=d$ and $d > 1/\eps$, and $\eps \in (0,c)$ for some constant $c$ that goes to $0$ (as $n$ and $d$ grow accordingly), then the size of any $\eps$-cover needs to be at least $\Omega((1/\eps)^{d^{1-o(1)}})$ where the $o(1)$ shrinks as d grows for $\log_{1/\eps}(d) \geq 1$.   
\end{corollary}

That is if $n=d$ and both $d$ and $1/\eps$ grow, then an $\eps$-cover requires nearly $1/\eps^d$ ranges, and how close it is to this bound depends on how much faster $d$ grows than $\log_2(1/\eps)$. 

\begin{proof} 
Let $\lambda \in (0, 1/2)$, $\eta \in (0.9, 1)$ and let $c = 1/a$ such that $1 - \eps \log_2(e/\eps) \geq \eta$ and $a$ is chosen such that for any $x > a$ the inequality $\log_2 x \leq \eta x^{\lambda}$ holds.  Let also $X = \{e_1, \ldots, e_d\}$ be the vertices of the $(d-1)$-simplex in $\R^d$, where $n = d = 1/\eps^k$ so $k= \log_{1/\eps}(d)$.  By changing the base $2$ in case (ii) of Theorem \ref{thm:lower-comb-eps-cover} to $1/\eps$ we obtain
\begin{equation} \label{eq: d-o(1)}
|Q| 
\geq 2^{d (1 - \eps \log_2(e/\eps))} 
= \Big(\frac{1}{\eps}\Big)^{d(1 - \eps \log_2(e/\eps)) \log_{1/\eps}(2)} 
\geq \Big(\frac{1}{\eps}\Big)^{\eta d/\log_2(1/\eps)} 
\geq \Big(\frac{1}{\eps}\Big)^{\eps^{\lambda}d}.
\end{equation}
Now, if $\lambda/k = o(1)$ hence $\lambda = o(k)$ and thus $(\frac{1}{\eps})^\lambda = (\frac{1}{\eps})^{o(k)}$ or equivalently $\eps^{\lambda} \geq d^{-o(1)}$.  Therefore, by \eqref{eq: d-o(1)}, $|Q| \geq (\frac{1}{\eps})^{\eps^{\lambda}d} \geq (\frac{1}{\eps})^{d^{1 - o(1)}}$. 
\end{proof}

\bibliographystyle{plain}
\bibliography{references}

\newpage
\appendix

\section{Appendix}

\subsection{Equivalence Between half-spaces, Fixed-Radius-Balls, and Any-Radius-Balls}
\label{app:H=frB=B}

Recall that $(X,\cH$), $(X, \cal{B})$ and $(X, {\cal{B}}_r)$ denote the half-space, ball and fixed-radius-$r$-ball range spaces respectively, where $X \subset \R^d$ is finite.  To be precise, $(X, \cal{B})$ shows all possible ranges defined by all possible balls with any radii, while $(X, {\cal{B}}_r)$ shows all possible ranges defined by all possible balls with radius $r$.  Our goal here is to show that these range spaces are roughly equivalent when considering large $d$. 

Consider a range space $(X, \cH)$.  Then for any range $R = X \cap h$ defined by a half-space $h$, we can choose a large enough radius $r_R$ and identify a radius-$r_R$ ball $B_{r_R}$ so $B_{r_R} \cap X = R = X \cap h$.  Indeed, for a sufficiently large radius $r$ for each $R \in (X, \cH)$, there exists a ball $B_r$ which corresponds to that range.  Therefore, if we choose $r = \max_{R \in (X, \cH)} r_R$ with appropriate centers for each ball, then $(X, \cH) \subset (X, {\cal{B}}_r)$.  
Thus any lower bound for $(X,\cH)$ also applies to $(X,{\cal{B}}_r)$.  

The inclusion $(X, {\cal{B}}_r) \subset (X, \cal{B})$ is trivial as ${\cal{B}}_r \subset \cal{B}$. 

Now let $R \in (X, \cal{B})$.  Then there is a ball $B_s(p)$ such that $R = B_s(p) \cap X$.  We consider the Veronese map $\Psi: \R^d \to \R^{d+1}$ by $x \mapsto (x, \|x\|^2)$.  We show that the $\Psi(R)$ can be obtained via the intersection of a half-space $h$ in $\R^{d+1}$.  Let $x = (x_1, \ldots, x_d) \in R$.  Then $\|x-p\|^2 \leq s^2$.  Rewriting this we get $\langle (-2p_1, \ldots, -2p_d, 1), (x_1, \ldots, x_d, \|x\|^2) \rangle \leq s^2 - \|p\|^2$, where $p = (p_1, \ldots, p_d)$.  This means that $\Psi(R) = \Psi(X) \cap h_{p, s}$, where $h_{p,s}$ is a half-space in $\R^{d+1}$ defined by $h_{p, s}(y) = \langle (-2p_1, \ldots, -2p_d, 1), (y_1, \ldots, y_d, y_{d+1}) \rangle + \|p\|^2 - s^2$.  Therefore, $(X, \cal{B})$ can have its corresponding in $(\Psi(X), \cH)$ where the Veronese map $\Psi$ lifts the dimension $d$ to $d+1$. 
Hence any lower bound for $(X, \cal{B})$ in $\R^d$ also applies to $(X,\cH)$ in $\R^{d+1}$.

\subsection{Relation Between $\eps$-KDE-Samples and $\eps$-Cover-Samples} \label{app:coreset}

The following lemma shows that an $\eps$-cover-sample is a $(1+c)\eps$-KDE-sample for any $c>0$.

\begin{lemma}\label{lem:coreset}
Let $S \subset X$ be such that for any $p,q \in \R^d$, $\big|d_{\symdiff}^X(p, q) - d_{\symdiff}^S(p, q)\big| \leq \eps$. Then $S$ is a $(1+c)\eps$-KDE-sample for any $c>0$ if the critical radius of $K$ is finite for any $\eps>0$. 
\end{lemma}
\begin{proof} 
Let $q\in \R^d$. Take a point $p \in \R^d$ at infinity (i.e. is very far from all data points). More precisely, take $p \in \R^d$ in such a way that $K(p, x)\leq \min\{c\eps/2, K(q, x)\}$ for all $x \in X$. Then $|K(q,x)-K(p,x)| = K(q,x)-K(p,x)$ and by setting $K_X(y) = \frac{1}{|X|}\sum_{x \in X} K(y, x)$ we have 
\[
\begin{array}{ll}
     \displaystyle |K_X(q) - K_S(q)| & 
     \leq |K_X(q) - K_X(p) + K_S(p) - K_S(q)| + |K_X(p) - K_S(p)|
     \vspace{1mm} \\ &
     \leq |K_X(q) - K_X(p) + K_S(p) - K_S(q)| + K_X(p) + K_S(p) \vspace{1mm} \\ &
    \leq \displaystyle\bigg|\frac{1}{|X|} \sum_{x\in X} |K(q,x) - K(p,x)| - \frac{1}{|S|} \sum_{s\in S} |K(q,s) - K(p,s)| \bigg| + c\eps \vspace{2mm} \\ & 
    = \big|d_{\symdiff}^X(p, q) - d_{\symdiff}^S(p, q)\big| + c\eps \vspace{2mm} \\ & 
    \leq (1+c)\eps. \hfill \qedhere
\end{array}  
\]
\end{proof}

Employing Lemma \ref{lem:coreset} along with Theorem \ref{thm:sym-coreset} we obtain the following corollary. 
\begin{corollary}\label{eps-sample}
Let $S$ be an $\eps$-sample for $(X,\cA)$, where $\cA$ is semi-linked to a kernel $K$,  where the critical radius of $K$ is finite for any $\eps>0$. Then $S$ is a $(1+c)\eps$-KDE-sample for any $c>0$. 
\end{corollary}

\subsection{More on $\eps$-KDE-Samples}
\label{app:KDE-samples}

Corollary \ref{cor: kernel-eps-sample-size} gives an input-size- and dimension-free upper bound on the $\eps$-KDE-samples.  We remark that if $K_1, K_2$ satisfy the conditions of Corollary \ref{cor: kernel-eps-sample-size}, $K_1+c K_2$ and $K_1K_2$ also meet the conditions too, for any constant $c>0$.  So, a wide variety of kernels work for Corollary \ref{cor: kernel-eps-sample-size}.  We introduce some non-characteristic kernels that do the job. 

Consider the truncated Gaussian kernel. That is, let $K(x,p) = e^{-\|x-p\|^2/\sigma^2}$ for $x$ in the super-level set $R_{p, \tau} = \{x : e^{-\|x-p\|^2/\sigma^2} \geq \tau\}$ for some $\tau \in (0,1)$, and $0$ elsewhere. Then the modified truncated Gaussian is $(e^{-\|x-p\|^2/\sigma^2} - \tau)/(1 - \tau)$ for $x \in R_{p, \tau}$ and $0$ elsewhere.  Note, the modified truncated Gaussian kernel is $(1, \sqrt{\ln(1/\eps)})$-standard $2$-simply computable like the Gaussian kernel.

\begin{corollary}\label{cor: kernel-eps-sample-size-1}
With probability $\geq 1-\delta$, a random sample from $X$ is an $\eps$-KDE-sample when $K$ is an Epanechnikov, quartic, triweight, triangle or modified truncated Gaussian kernel and the sample size is $O(\frac{1}{\eps^6} \log^2 \frac{1}{\eps \delta})$.
\end{corollary}

\subsection{Constructing $\eps$-Cover-Samples from $\eps$-Samples} \label{app:coreset-semi-linked}

In the following theorem we restate and prove Theorem \ref{thm:sym-coreset}. As we mentioned before Theorem \ref{thm:sym-coreset}, the proof technique mostly is borrowed from  Joshi \emph{et al.} \cite{joshi2011comparing}.

\begin{theorem}[Restatement of Theorem \ref{thm:sym-coreset}]\label{thm:sym-coreset-restatement}
Let $S$ be an $\eps$-sample for $(X,\cA)$, where $\cA$ is semi-linked to a kernel $K$, where $K\leq 1$. Then $S$ is an $\eps$-cover-sample for $X$. 
\end{theorem}
\begin{proof}
We need to show that for any $p,q \in \R^d$, $\big|d_{\symdiff}^X(p, q) - d_{\symdiff}^S(p, q)\big| \leq \eps$. 

Suppose $X=\{x_1, \ldots, x_n\}$, $S=\{s_1, \ldots, s_m\}$ and $k=n/m$, where without loss of generality we can assume that $k$ is an integer (otherwise we can work with fractional assignments). Moreover, for the sake of convenience let $E = d_{\symdiff}^X(p, q) - d_{\symdiff}^S(p, q)$. In order to get the desired inequality, we design two different partitions for $X$ and show that $E \leq \eps$ and $E \geq -\eps$. Given $p$ and $q$, again for simplicity let $f(x)=|K(p, x) - K(q,x)|$ (note that $f(x) \leq 1$). 

{\bf Undercounts.} For the first partition we sort $X$ and $S$ in decreasing way by their $f$ value, i.e. 
\[
f(x_1) \geq f(x_2) \geq \cdots \geq f(x_n) \quad \text{and} \quad f(s_1) \geq f(s_2) \geq \cdots \geq f(s_m).
\]
Without loss of generality one may assume that $f(s_1) > f(s_2) > \cdots > f(s_m)$ by a tiny perturbation of $S$. Then any semi-super-level set containing $x_i$ ($s_i$ respectively) will also contain all $x_j$ ($s_j$ respectively) for $j<i$. Then we consider $2m$ (possibly empty) sets $\{P_1, \ldots, P_m\} \cup \{Q_1, \ldots, Q_m\}$ using the sorted order by $f$. Starting with $x_1$ (the point with highest $f$ value) we place points in $P_j$ or $Q_j$ following their sorted order. Starting at $i=j=1$, we place $x_i$ in $Q_j$ as long as $f(x_i) > f(s_j)$ (this can be empty). 
Then we place the next $k$ points into $P_j$. After these $k$ points we start by $Q_{j+1}$ and place points in $Q_{j+1}$ as long as $f(x_i) > f(s_{j+1})$. Then we put the next $k$ points of $X$ into $P_{i+1}$. We continue this process until all of $X$ has been placed in some set. Let $t \leq m$ be the index of last set $P_j$ such that $|P_j|=k$. Then $|P_{t+1}| <k$ and $P_j=Q_j=\emptyset$ for all $j > t+1$. We also observe that for all $x_i \in P_j$ (for $j \leq t$) we have $f(s_j) \geq f(x_i)$ and so $k f(s_j) \geq \sum_{x_i \in P_j} f(x_i)$ or equivalently, $\frac{1}{m}f(s_j) \geq \frac{1}{n} \sum_{x_i \in P_j} f(x_i)$. We can now bound the undercounts as 
\begin{equation} \label{eq4}
\begin{array}{ll}
E \!\!\! & 
= \displaystyle \frac{1}{n} \sum_{i=1}^n f(x_i) - \frac{1}{m} \sum_{j=1}^m f(s_j) = \sum_{j=1}^m \Big(\frac{1}{n} \sum_{x_i\in P_j}f(x_i) + \frac{1}{n} \sum_{x_i\in Q_j}f(x_i)\Big) - \frac{1}{m} \sum_{j=1}^m f(s_j) \\ &
= \displaystyle \sum_{j=1}^m \underbrace{\Big(\frac{1}{n} \sum_{x_i\in P_j}f(x_i) - \frac{1}{m} f(s_j)\Big)}_{\leq 0} + \sum_{j=1}^m \Big(\frac{1}{n} \sum_{x_i\in Q_j}f(x_i)\Big) \leq \sum_{j=1}^{t+1} \Big(\frac{1}{n} \sum_{x_i\in Q_j}f(x_i)\Big) \\ & 
\leq \displaystyle \frac{1}{n} \sum_{j=1}^{t+1} |Q_j| = \frac{1}{n} \sum_{j=1}^{t+1} |Q_j \cap A|,
\end{array}
\end{equation}
where $A$ is the semi-super-level set $A = \{x \in \R^d: f(x) \geq \tau\} \in \cA$ with $\tau = f(x_l)$, where $l$ is the largest index such that $f(x_l) > f(s_{t+1})$.  Then $A$ contains $s_t$ but not $s_{t+1}$.  Therefore, $s_j \in A$ for $j \leq t$ and $s_j \notin A$ for $j\geq t+1$, and so $|P_j \cap A| = k$ for $j \leq t$ and $|P_j \cap A| = 0$ for $j \geq t+1$. Since $S$ is an $\eps$-sample for $(X, \cA)$, then 
\begin{equation} \label{eq5}
\sum_{j=1}^{t+1} |Q_j \cap A| = \bigg(\sum_{j=1}^{t+1} |Q_j\cap A| + \sum_{j=1}^t |P_j \cap A|\bigg) - k|S \cap A| = |X \cap A| -  k|S \cap A| \leq n \eps.
\end{equation}
Therefore, using \eqref{eq4} and \eqref{eq5} we can write $E \leq \eps$. 

{\bf Overcounts.} For the second partition we do overcounts analysis similar to undercounts.  In this partitioning we sort $X$ and $S$ in increasing way by their $f$ value, i.e. 
\[
f(x_1) \leq f(x_2) \leq \cdots \leq f(x_n) \quad \text{and} \quad f(s_1) \leq f(s_2) \leq \cdots \leq f(s_m).
\]
Again without loss of generality we assume that $f(s_1) < f(s_2) < \cdots < f(s_m)$. Then we consider $2m$ (possibly empty) sets $\{P_1, \ldots, P_m\} \cup \{Q_1, \ldots, Q_m\}$ using the sorted order by $f$. Starting with $x_1$ (the point with lowest $f$ value) we place points in $P_j$ or $Q_j$ following their sorted order. Starting at $i=j=1$, we place $x_i$ in $Q_j$ as long as $f(x_i) < f(s_j)$ (this may be empty). Then we place the next $k$ points $x_i$ into $P_j$. After $k$ points are placed in $P_j$, we begin with $Q_{j+1}$ until all of $X$ has been placed in some set. Let $t \leq m$ be the index of last set $P_j$ such that $|P_j|=k$. Then $|P_{t+1}| <k$ and $P_j=Q_j=\emptyset$ for all $j > t+1$. We also observe that for all $x_i \in P_j$ (for $j \leq t$), $f(s_j) \leq f(x_i)$, and thus $\frac{1}{m}f(s_j) \leq \frac{1}{n} \sum_{x_i \in P_j} f(x_i)$. We can now bound the overcounts as 
\begin{equation} \label{eq6}
\begin{array}{ll}
E \!\!\! & 
= \displaystyle \frac{1}{n} \sum_{i=1}^n f(x_i) - \frac{1}{m} \sum_{j=1}^m f(s_j) = \sum_{j=1}^m \Big(\frac{1}{n} \sum_{x_i\in P_j}f(x_i) + \frac{1}{n} \sum_{x_i\in Q_j}f(x_i)\Big) - \frac{1}{m} \sum_{j=1}^m f(s_j) \\ &
= \displaystyle \sum_{j=1}^t \underbrace{\Big(\frac{1}{n} \sum_{x_i\in P_j}f(x_i) - \frac{1}{m} f(s_j)\Big)}_{\geq 0} + \sum_{j=t+1}^m \Big(\frac{1}{n} \sum_{x_i\in P_j}f(x_i) - \frac{1}{m} f(s_j)\Big) + \underbrace{\sum_{j=1}^m \Big(\frac{1}{n} \sum_{x_i\in Q_j}f(x_i)\Big) }_{\geq 0} \\ & 
\geq \displaystyle \sum_{j=t+1}^m \Big(\frac{1}{n} \sum_{x_i\in P_j}f(x_i) - \frac{1}{m} f(s_j)\Big) = \Big(\frac{1}{n} \sum_{x_i\in P_{t+1}}f(x_i) - \frac{1}{m} f(s_{t+1})\Big) - \frac{1}{m} \sum_{j=t+2}^m f(s_j) \\ & 
\geq \displaystyle \frac{1}{n} \sum_{x_i\in P_{t+1}}f(x_i) - \frac{1}{m} f(s_{t+1}) - \frac{(m-t-1)}{m}.
\end{array}
\end{equation}
Now let $A \in \cA$ be a semi-super-level set containing no point from $\cup_{j=1}^m Q_j$. For instance, $A$ can be $R_{p,q, f(s_u)}$, where $u$ is the largest index such that $Q_u \neq \emptyset$ (note $u \leq t+1$). Then $X \cap A = P_u \cup \cdots \cup P_{t+1}$ and $S \cap A = \{s_u, \ldots, s_m\}$, and so $|X\cap A| = (t+1-u)k+|P_{t+1}|$ and $|S \cap A| = m-u+1$. Hence, because $S$ is an $\eps$-sample for $(X, \cA)$, we have 
\[
\begin{array}{ll}
\eps & \!\!\!  
\displaystyle \geq \frac{1}{m} |S \cap A| - \frac{1}{n} |X \cap A| = \frac{m-u+1}{m} - \frac{(t+1-u)k+|P_{t+1}|}{n} \vspace{2mm} \\ &
\displaystyle =  \frac{m-u+1}{m} - \frac{(t+1-u)}{m} - \frac{|P_{t+1}|}{n} = \frac{m-t}{m} - \frac{|P_{t+1}|}{n}, 
\end{array}
\]
which implies $m-t-1 \leq m \eps + \frac{m}{n} |P_{t+1}| - 1$. Let $f$ attain its minimum on $P_{t+1}$ at $x_i \in P_{t+1}$, so $f(x_i) \geq f(s_{t+1})$. Then by applying \eqref{eq6} we can infer 
\[
\begin{array}{ll}
E \!\!\!  &
\displaystyle \geq \frac{1}{n} \sum_{x_i\in P_{t+1}}f(x_i) - \frac{1}{m} f(s_{t+1}) - \frac{(m \eps + \frac{m}{n} |P_{t+1}| - 1)}{m} \\ &
\displaystyle = - \eps + \Big(\frac{k- |P_{t+1}|}{n} \Big) - \frac{1}{n} \Big( \sum_{x_i \in P_{t+1}} f(x_i) - k f(s_{t+1}) \Big) \\ &
\displaystyle \geq - \eps + \Big(\frac{k- |P_{t+1}|}{n} \Big) - \Big(\frac{k- |P_{t+1}|}{n} \Big) f(x_i) \geq  - \eps. 
\end{array} 
\]
\end{proof}

\subsection{Missing Proof Elements for Lower Bound} \label{sec:lower bound}

\begin{lemma} [Restatement of Lemma \ref{lem: sphere intersection}]\label{lem: sphere intersection-proof}
Let $S_1, \ldots, S_d$ be $d$-spheres in $\R^d$, where $S_k$ is centered at $e_k$ with radius $r_k \geq 1$ ($e_k$ is the standard $k$-th basis vector in $\R^d$). Assume that there are $r \geq 1$ and $\delta \in (0,1/d)$ such that $|r_k^2 - r^2| < \delta$ for $k=1, \ldots, d$. Then $|\bigcap_{k=1}^d S_k| = 2$.
\end{lemma}
\begin{proof}
We are looking for two points like $x=(x_1, \ldots,x_d) \in \mathbb{R}^d$ such that $\|x-e_k\|^2 = r_k^2$ for $k=1, \ldots, d$. Writing each equation and gathering similar terms yields
\begin{equation}\label{eq1s}
\textstyle x_k = \frac{1}{2} (1+\|x\|^2-r_k^2), \quad k=1, \ldots, d.
\end{equation}
If the system of equations \eqref{eq1s} has a solution, then $y = \|x\|^2$ would be one of the solutions to the quadratic equation
\begin{equation}\label{eq2s}
y = \frac{1}{4} \sum_{k=1}^d (1+y-r_k^2)^2 = \frac{d}{4} y^2 + \frac{1}{2} \sum_{k=1}^d(1-r_k^2)y + \frac{1}{4} \sum_{k=1}^d(1-r_k^2)^2,
\end{equation}
or equivalently,
\begin{equation}\label{eq3s}
p(y) = \frac{d}{4} y^2 + \Big(\frac{1}{2} \sum_{k=1}^d(1-r_k^2) - 1\Big)y + \frac{1}{4} \sum_{k=1}^d(1-r_k^2)^2=0.
\end{equation}
Now we consider the discriminant to obtain the criteria to guarantee  existence of 2 distinct solutions to the quadratic equation \eqref{eq3s}:
\[
\Delta = \Big( \frac{1}{2} \sum_{k=1}^d(1-r_k^2) -1 \Big)^2 - \frac{d}{4} \sum_{k=1}^d(1-r_k^2)^2 = \frac{1}{4} \Big(\sum_{k=1}^d(1-r_k^2)\Big)^2  + 1 + \sum_{k=1}^d(r_k^2-1)- \frac{d}{4} \sum_{k=1}^d(1-r_k^2)^2.
\]
By setting $B = (b_1, \ldots, b_d)$, where $b_i=r_i^2-1$ and $b=r^2-1$, and using $|b_k-b|<\delta$ we get 
\begin{equation}\label{eq7s}
\begin{array}{ll}
4 \Delta \!\!\! & = \|B\|_1^2 + 4 + 4 \|B\|_1 - d \|B\|_2^2 \geq d^2 (b-\delta)^2 - d^2 (b+\delta)^2 + 4d (b-\delta) + 4  \vspace{2mm} \\ & 
= - 4d^2 \delta b + 4 db - 4d \delta +4 = 4 (db + 1)(1-d \delta).
\end{array}
\end{equation}
The condition $\delta < 1/d$ implies $\Delta>0$, which shows that the quadratic equation \eqref{eq3s} has two roots, say $y_1, y_2$. Moreover, these two roots are positive. (Notice the roots of the quadratic equation $y^2-by+c=0$ are real and positive if and only if $b>0$ and $b^2 \geq 4c>0$.) Substituting $y_1, y_2$ in \eqref{eq1s} we obtain two points in the intersection of spheres $S_1, \ldots, S_d$ as
\[
\textstyle x_k = \frac{1}{2} (1+y_1-r_k^2), \quad k=1, \ldots, d,  \quad \text{and} \quad
x_k = \frac{1}{2} (1+y_2-r_k^2), \quad k=1, \ldots, d.
\]
Therefore, $|\bigcap_{k=1}^d S_k| = 2$. 
\end{proof}

\subsection{Simple Computability of Triangle Kernel} \label{app:VC-tri}

The following theorem shows that, like the Gaussian and Epanechnikov kernels, triangular kernels are $2$-simply computable.

\begin{theorem}\label{thm:vc-dim: triangular}
The triangular kernel $K(x,y) = \max(0, 1 - \|x-y\|)$ is $2$-simply computable. 
\end{theorem}
\begin{proof}
Let $p,q,x \in \R^d$, $\tau \in \R^+$ and consider the inequality $|K(p, x) - K(q, x)| \geq \tau$. There are three cases: 
\begin{enumerate}
    \item $\|p-x\| < 1$ and $\|q-x\| \geq 1$. Then verifying $|K(p, x) - K(q, x)| \geq \tau$ is equivalent to verifying $\|x-p\| \leq 1 - \tau$.
    
    \item $\|p-x\| \geq 1$ and $\|q-x\| < 1$. Then verifying $|K(p, x) - K(q, x)| \geq \tau$ is equivalent to verifying $\|x-p\| > 1 - \tau$.
    
    \item $\|p-x\|<1$ and $\|q-x\|<1$. Then writing down the inequality $|K(p, x) - K(q, x)| \geq \tau$, we get
    \[
    \textstyle \sqrt{\sum_{i=1}^d (x_i-p_i)^2} \geq \tau + \sqrt{\sum_{i=1}^d (x_i-q_i)^2} \quad \text{or} \quad \textstyle \sqrt{\sum_{i=1}^d (x_i-q_i)^2} \geq \tau + \sqrt{\sum_{i=1}^d (x_i-p_i)^2}.
    \]
    Consider the left hand side inequality (the other comes by symmetry).  Squaring both sides and simplifying the equation we obtain the following equation:
    \begin{equation} \label{eq: hyperbola}
    \Big[- \tau^2 + 2 \sum_{i=1}^d ((q_i-p_i)x_i + p_i^2 -q_i^2) \Big]^2 \geq 4 \tau^2 \sum_{i=1}^d (x_i-q_i)^2,
    \end{equation}
where $x=(x_1, \ldots, x_d)$, $p=(p_1, \ldots, p_d)$ and $q=(q_1, \ldots, q_d)$. 
\end{enumerate}
The cases 1 and 2 can be computed with $O(d)$ arithmetic operations or jumps conditions. The equation \eqref{eq: hyperbola} shows that Case 3 can also be computed in $O(d)$ time applying the same operations. Therefore, $K$ is $2$-simply computable.  
\end{proof}

\section{Terminal JL} \label{app:TJL}

The algorithm to compute terminal JL, taken almost verbatim from \cite{IP2022}, is as follows: 

\begin{lstlisting}[caption={Terminal JL.},label=list:alg:eps-net, abovecaptionskip=-\medskipamount, mathescape]
Input: $\eps \in (0,1)$, $X \subset \R^d$, $|X|=n$, $Q \subset \R^d \setminus X$ and $|Q| = k$.
For any $x \in X$ set $f(x) = (\Pi x, 0)$, where $\Pi$ is a JL map (explained below).
For $q \in Q$:
    (1) Compute $x_{NN} = \argmin_{x \in X} \|x-q\|$,
    (2) Solve the following constrained optimization problem:
        Minimize     $h_{q, x_{NN}}(z) = \|z\|^2 + 2\langle \Pi(q - x_{NN}), z \rangle$ 
        Subject to   $\|z\|^2 \leq \|q-x_{NN}\|$
                     $|\langle z, \Pi(x-x_{NN})\rangle - \langle q-x_{NN}, x-x_{NN}\rangle | \leq \eps \|q-x_{NN}\| \|x-x_{NN}\|$  
                     ($\forall x \in X$),
    (3) Let $q'$ be the solution to the minimization problem in (2). Set 
        $f(q) = (\Pi x_{NN} + q', \sqrt{\|q-x_{NN}\|^2 -\|q'\|^2}).$ 
Return $f$.
\end{lstlisting}

\subparagraph*{Run time analysis.}

First we need to construct a JL map, i.e. a random matrix $\Phi \in \R^{m\times d}$ with entries from normal distribution $N(0,1)$ normalized by $1/\sqrt{m}$, say $\Pi = 1/\sqrt{m} \Phi$, where $m = O(\log(n)/\eps^2)$. So, we can consider $O(dm)$ as its run time. Calculating $\Pi x$ for any $x \in X$ needs $O(md)$ times and so for the whole $X$ we need $O(nmd)$ time. Therefore, the run time for embedding $X$, i.e. computing $\Pi X$, is $O(dm + nmd) = O(d n\log(n)/\eps^2))$.

Now we need to compute the run time for embedding a single $q \in \R^d \setminus X$ in Terminal JL algorithm. 
Step 1 for finding the nearest neighbor $x_{NN}$ of $q$ needs at most $O(nd)$ time.

The optimization problem for finding $f(q)$ is a semidefinite programming since we can rewrite it as
\[
\begin{array}{rl}
     \text{\bf Minimize}    & t \\
     \text{\bf Subject to}  & \langle z, z\rangle + \langle 2 \Pi(q - x_{NN}), z \rangle \leq t \vspace{1mm} \\
                        & \langle z, z \rangle \leq \|q-x_{NN}\| \vspace{1mm} \\
                        & \langle z, \Pi(x-x_{NN})\rangle \leq \eps \|q-x_{NN}\| \|x-x_{NN}\| + \langle q-x_{NN}, x-x_{NN}\rangle \ \ \forall x \in X \vspace{1mm} \\
                        & \langle z, \Pi(x_{NN}-x)\rangle \leq -\eps \|q-x_{NN}\| \|x-x_{NN}\| - \langle q-x_{NN}, x-x_{NN}\rangle \ \ \forall x \in X
\end{array}
\]
So, we have $2n+2$ constraints on $z \in \R^d$. Therefore, according to the literature, the running time for computing $f(u)$, given $x_{NN}$, is $O(\sqrt{d}(n^3+d^3)\log(1/\eps))$, see \cite{semidef2020}. 
Therefore, $O(d n\log(n)/\eps^2 + \sqrt{d}(n^3+d^3)\log(1/\eps))$ can be a (non-optimal) upper bound on the running time of embedding $X \cup \{q\}$ via terminal JL.

\section{Rademacher Complexity of Unit Ball in RKHS}

The following theorem is well known but we could not find it in the literature to cite.  So, we present the theorem here and include the proof from \cite{lecture} for completeness.  According to  Definition 3.1 of \cite{MRT2018} The empirical  Rademacher complexity of a function class $\cal F$ (functions mapping $X$ to $[a,b]$) with respect to the sample $S = \{s_1, \ldots, s_m\}$ from $X$ is defined as 
\[
\hat{\mathfrak{R}}_S(\mathcal{F}) = \E_{\sigma} \bigg[\sup_{f \in \mathcal{F}} \frac{1}{m} \sum_{i=1}^m \sigma_i f(s_i) \bigg],
\]
where $\sigma = (\sigma_1, \ldots, \sigma_n)$ and $\sigma_i$'s are independent uniform random variables from $\{-1,1\}$. 

\begin{theorem}\label{thm:Rademacher-RKHS}
Let $K$ be a positive definite bounded kernel with $\sup_{x} \sqrt{K(x,x)} = B$ and let $\cH$ be its RKHS.  Then for any sample $S = \{s_1, \ldots, s_m\}$, 
$\hat{\mathfrak R}_S(B_1^\cH(0)) \leq \frac{B}{\sqrt{m}}$, where $B_1^\cH(0) = \{f \in \cH : \|f\|_{\cH} \leq 1\}$.
\end{theorem}
\begin{proof}
Fix $S = \{s_1, \ldots, s_m\}$.  Then 
\[
\begin{array}{ll}
    \hat{\mathfrak R}_S(B_1^\cH(0)) & 
    = \displaystyle \E_{\sigma} \bigg[\sup_{f \in B_1^\cH(0)} \frac{1}{m} \sum_{i=1}^m \sigma_i f(s_i) \bigg]  
    = \displaystyle \frac{1}{m} \E_{\sigma} \bigg[\sup_{f \in B_1^\cH(0)} \sum_{i=1}^m \sigma_i \langle f, K(\cdot, s_i) \rangle \bigg] \vspace{1mm} \\ &
    = \displaystyle \frac{1}{m}  \E_{\sigma} \bigg[\sup_{f \in B_1^\cH(0)} \Big\langle f, \sum_{i=1}^m \sigma_i K(\cdot, s_i) \Big\rangle \bigg] \vspace{1mm} \\ &
    = \displaystyle \frac{1}{m} \E_{\sigma} \bigg[\Big\langle \frac{\sum_{i=1}^m \sigma_i K(\cdot, s_i)}{\|\sum_{i=1}^m \sigma_i K(\cdot, s_i)\|_\cH}, \sum_{i=1}^m \sigma_i K(\cdot, s_i) \Big\rangle \bigg] \vspace{1mm} \\ &
    = \displaystyle \frac{1}{m} \E_{\sigma} \bigg[\Big\|\sum_{i=1}^m \sigma_i K(\cdot, s_i)\Big\|_\cH \bigg] 
    = \displaystyle \frac{1}{m} \E_{\sigma} \left[\sqrt{\Big\|\sum_{i=1}^m \sigma_i K(\cdot, s_i)\Big\|_\cH^2} \right] \vspace{1mm} \\ &
    \leq \displaystyle \frac{1}{m} \sqrt{\E_{\sigma} \Big\|\sum_{i=1}^m \sigma_i K(\cdot, s_i)\Big\|_\cH^2} 
    = \displaystyle \frac{1}{m} \sqrt{\sum_{i=1}^m \|K(\cdot, s_i)\|_\cH^2} \vspace{1mm} \\ &
    = \displaystyle \frac{1}{m} \sqrt{\sum_{i=1}^m K(s_i, s_i)} \leq \frac{1}{m} \sqrt{m B^2} = \frac{B}{\sqrt{m}}.
\end{array}
\]
We used Jensen’s inequality, reproducing property, $\E_{\sigma}[\sigma_i \sigma_j] = 0$ for $i \neq j$, and the fact that a bounded linear functional obtains its norm in its normalized representer according to the Riesz representation theorem. 
\end{proof}
\end{document}